\documentclass[12equiped]{article}
\usepackage{graphicx}
\usepackage{amssymb,amsmath}
\usepackage{hyperref}
\usepackage{placeins}
\input epsf.tex

\title{Moduli space of Conformal Field Theories and non-commutative Riemannian geometry}

\author{Yan Soibelman}

\begin{document}
\maketitle
\newtheorem{thm}{Theorem}[subsection]
\newtheorem{defn}[thm]{Definition}
\newtheorem{lmm}[thm]{Lemma}
\newtheorem{rmk}[thm]{Remark}
\newtheorem{prp}[thm]{Proposition}
\newtheorem{conj}[thm]{Conjecture}
\newtheorem{exa}[thm]{Example}
\newtheorem{cor}[thm]{Corollary}
\newtheorem{que}[thm]{Question}
\newtheorem{ack}{Acknowledgements}
\newcommand{\C}{{\bf C}}
\newcommand{\K}{{\bf k}}
\newcommand{\R}{{\bf R}}
\newcommand{\N}{{\bf N}}
\newcommand{\Z}{{\bf Z}}
\newcommand{\Q}{{\bf Q}}
\newcommand{\G}{\Gamma}
\newcommand{\A}{A_{\infty}}
\newcommand{\g}{\mathfrak g}
\newcommand{\ihom}{\underline{\Hom}}
\newcommand{\epi}{\twoheadrightarrow}
\newcommand{\mono}{\hookrightarrow}
\newcommand{\ra}{\longrightarrow}
\newcommand\uhom{{\underline{Hom}}}
\renewcommand\O{{\cal O}}
\newcommand\nca{nc{\bf A}^{0|1}}
\newcommand{\lan}{\langle}
\newcommand{\ran}{\rangle}
\newcommand{\epp}{\varepsilon}

\tableofcontents

\section{Introduction}

\subsection{Motivations}\label{motivations}
This paper is  a slightly revised  version of [So1]. I have decided to submit it here for two reasons. First, ideas of loc.cit.  have attracted attention recently (see e.g. [OW]), while the paper itself is not easy to get. Second, I have added as an appendix  a draft of our 20-year old unfinished work with Maxim Kontsevich devoted to the deformation theory of quantum field theories (QFTs for short) as well as speculations about QFTs on compact metric spaces. Morally these topics are related to the  rest of the paper. I have not done a substantial revision of [So1]. This concerns  the list of references as well. I apologize for omissions of many related papers published since then.

Main idea of [So1] (and hence of this paper) is the {\it analogy between the moduli space of Conformal Field Theories (CFTs for short) and
the space of isomorphism classes of compact metric-measure spaces equipped with the measured Gromov-Hausdorff topology}.

Algebro-geometric approach to the concept of  moduli space as to the space representing the functor of ``isomorphism classes of families" is not very useful for Riemannian manifolds\footnote{Same can be said about many other ``functorial" concepts.} Maybe this is the reason why the moduli space of CFTs has not been defined in the functorial way despite of several attempts, starting with the one of Graeme Segal himself.

On the other hand, the ``moduli space of compact Riemannian manifolds" naively understood as the set of isometry classes of compact
Riemannian manifolds carries some  natural Hausdorff topologies (e.g. the one coming from Gromov-Hausdorff metric). Therefore one can ``compactify\rq\rq{}
it in the larger space consisting of isometry classes of compact metric spaces. It is well-known that
some differential-geometric structures of Riemannian geometry admit generalizations to this compactification. As an example we mention the notion of sectional curvature
extended to Alexandrov spaces or the property to have non-negative Ricci curvature
extended to compact metric-measure spaces (see [LV], [St]).

It is natural to ask whether this philosophy can be generalized
to the case of properly defined {\it non-commutative} Riemannian manifolds. The corresponding ``moduli space of non-commutative Riemannian manifolds\rq\rq{} should contain moduli space of CFTs inside, while the boundary it should contain certain degenerations of CFTs which do not have to be CFTs. To my knowledge, this program is still in its infancy, twenty years after [So1].

In this paper we propose an approach to non-commutative Riemannian geometry  motivated by Segal's axioms of the unitary Conformal Field Theory (see [Seg]) as well as a version of non-commutative Riemannian geometry
developed by A. Connes in  [Co1]. Many structures of Connes's approach (e.g. spectral triples, see [Co1], [CoMar])
are closely related to the structures considered in present paper. 

We should warn the reader that the paper does not  contain results which can be called ``new" by mathematical or physical standards. Mostly it  is  a review and discussion of various existing concepts united by the author's wish to see their analogs in the framework of non-commutative (a.k.a ``quantum") geometry. In which sense these new analogs are really non-commutative is a different question.

\subsection{Collapsing CFTs from metric point of view}\label{collapsing CFTs and Ricci curvature}

One of our goals is to define a quantum Riemannian
manifold (or, more generally, ``Riemannian space") by a set of axioms similar to Segal's axioms of a unitary CFT.
Set of ``isometry classes" (i.e. the moduli space)  should be treated similarly to the set of isometry classes
of compact Riemannian manifolds with restrictions on the diameter
and Ricci curvature. The central charge plays a role of the dimension
of the space, and  the ``spectral gap" for the Virasoro operator $L_0+\overline{L}_0$
plays a role of the (square root of the inverse to) diameter. In the spirit of Gromov, Cheeger, Colding, Fukaya and others
we would like to compactify the ``moduli space" of such objects by their
``Gromov-Hausdorff degenerations".

As we have already mentioned above, we would like to treat the moduli space of CFTs (or,more generally, QFTs)
as an object of metric geometry rather than the one of the algebraic geometry.
We used this philosophy in [KoSo1], where
the concept of collapsing family of unitary CFTs was introduced with the aim
to explain Mirror Symmetry. Main idea of [KoSo1] is that the ``moduli
space of CFTs with the bounded central charge and the spectral gap\footnote
{i.e. the minimal positive eigenvalue of the Virasoro operator $L_0+\overline{L}_0$}
bounded from below by a non-negative constant", is precompact in some natural topology.
The topology itself was not specified in [KoSo1].  Since the notion of collapse depends
on the spectral properties of the Virasoro operator $L_0+\overline{L}_0$, one should use
the topology which gives continuity of the spectral data, e.g. measured Gromov-Hausdorff topology.
It was argued in loc. cit. that if
the spectral
gap approaches to zero, then the  collapsing family of unitary CFTs  gives rise to a
topological space, which contains a dense open Riemannian manifold with non-negative
Ricci curvature. The restriction on the Ricci curvature follows from the unitarity of the theory.
This can be encoded into a slogan:

{\it Collapsing unitary two-dimensional CFTs give rise to Riemannian manifolds (possibly singular) with non-negative Ricci curvature.}

From this point of view the
geometry of possibly singular Riemannian manifolds with non-negative Ricci curvature
should be thought of as a limit of the (quantum) geometry  of certain Quantum Field Theories (e.g. two-dimensional unitary CFTs). Riemannian manifolds themselves
appear as target spaces of non-linear sigma models. As such non-linear sigma-models provide a partial compactification of the moduli space of two-dimensional unitary CFTs.

\subsection{Quantum Riemannian spaces via Segal's axioms}\label{quantum Riemann spaces}

From the point of view of Segal's axioms, the  geometry
which underlies a collapsing family of CFTs is the geometry of 2-dimensional compact oriented Riemannian
manifolds degenerating into metrized graphs. The algebra of the CFT is encoded into the operator product
expansion (OPE). Its collapse gives rise to a commutative algebra $A$.
The (rescaled) operator
$L_0+\overline{L}_0$ collapses into the second order differential operator  on $A$. The rest of the conformal group does not survive. Hence the family of CFTs collapses into a QFT. The latter, according to [KoSo1],
should be thought of as a Gromov-Hausdorff limit of the former.

The above considerations suggest the following working definitions in the framework of quantum geometry.
A quantum compact Riemannian
space (more precisely, $2$-space, if we want to stress that surfaces are $2$-dimensional)
is defined by the following data: a separable complex Hilbert space $H$,
an operator $S({\Sigma}): H^{\otimes n}\to H^{\otimes m}$ called the amplitude of $\Sigma$ which is given
for each compact Riemannian $2$-dimensional oriented manifold $\Sigma$ with
$n$ marked ``input" circles and $m$ marked ``output" circles.  The kernel of $S({\Sigma})$
is the tensor $K_{\Sigma}\in H^{\otimes m}\otimes (H^{\otimes n})^{\ast}=
Hom(H^{\otimes n})^{\ast}\otimes H^{\otimes m},\C)$ called the correlator.
In the same vein a quantum compact Riemannian $1$-space is given by the similar
data assigned to metrized graphs with marked input and output vertices.
Natural gluing axioms should be satisfied in both cases, as well
as  continuity of the data with respect to some natural topology.
In particular we allow degenerations of  Riemannian $2$-spaces into  Riemannian
$1$-spaces. At the level of geometry this means that
metrized graphs are limits of
compact oriented Riemannian $2$-dimensional manifolds with boundary. At the level of algebra
all ``algebraic data" (e.g. spaces of states, operators) associated with  graphs are  limits of
the corresponding data for the surfaces. The notion of limit should have ``geometric" and ``algebraic"
counterparts. Geometrically it can be the measured Gromov-Hausdorff limit, while for Hilbert spaces
it can be any notion of limit which respects continuity of the spectral data of the associated
positive self-adjoint operators. We are going to review several possibilities in the main
body of the paper.

Also, we can (and should) relax the condition that $H$
is a Hilbert space, since the limit of Hilbert spaces can be a locally convex vector space
of more general type (e.g. a nuclear space). In order to obtain a ``commutative" Riemannian
geometry we require that the amplitude operator (or correlator) associated with a surface or a graph  is invariant with respect to the
(separate) permutations
of inputs and outputs.

If we accept the above working definitions of  quantum Riemannian spaces, then many natural questions arise, in particular:

a) What is the dimension of a quantum Riemannian space?

b) What is the diameter?

c) Which quantum Riemannian spaces should be called
manifolds?

d) What
are various curvature tensors, e.g. Ricci curvature?

When we speak about quantum
Riemannian $1$-spaces, Riemannian $2$-spaces or, more generally, Riemannian $d$-spaces, the number $d$ corresponds
to the dimension of the {\it world-sheet}, not the {\it space-time}. In particular, one can associate a Riemannian $1$-space with
a compact Riemannian manifold of any dimension. Having in mind possible relationship with
CFT we should allow the number $d$ to be non-integer (``central charge").

By analogy with the commutative Riemannian geometry one can ask about the structure of the
``moduli space" of quantum Riemannian manifolds rigidified by some geometric
data. In particular, we can ask about the ``space of isometry classes" of quantum
compact Riemannian $d$-spaces equipped with a non-commutative version of the Gromov-Hausdorff topology.
Then one can ask about analogs of classical precompactness and compactness theorems,
e.g. those
which claim precompactness of
the ``moduli space" of Riemannian manifolds having  fixed dimension,
diameter bounded from above and the Ricci curvature bounded from below (see e.g. [Gro1]).  

Introducing
a non-commutative analog of measure, one can ask about non-commutative analogs of theorems due to Cheeger, Colding, Fukaya
and others for the class
of {\it metric-measure spaces}, i.e. metric spaces equipped with a Borel probability
a measure. To compare with the case of  unitary CFTs we remark that
the space of states of a unitary CFT plays a role of $L_2$-space (of the loop space of a manifold),
with the vacuum expectation value playing a role of the measure.
The unitarity condition (``reflection positivity" in the language of Euclidean Quantum Field Theory) turns out to be an analog of the non-negativeness of the
Ricci curvature. Normalized boundary states should correspond to probability measures.

\

The above discussion motivates the idea to
treat unitary CFTs and their degenerations
as ``quantum metric-measure spaces with bounded diameter and
non-negative Ricci curvature".
One  expects that this moduli space is
precompact and complete in the natural topology.\footnote{More precisely, non-commutativity arises from CFTs with boundary conditions. Our point of view  differs from [RW] where a unitary
CFT already gives rise to a non-commutative space.
We prefer to axiomatize the structure arising from the full space of states rather
than from the subspace of invariants with respect to a $W$-algebra.} One hopes
for a similar picture for QFTs which live on the space-time of a more general type than just a manifold.
Leaving aside possible physical applications, one can ask (motivated by metric geometry)
``what is a QFT with the space-time, which is a compact metric-measure space ?" Although the
answer is not known, see Appendix for some  ideas in this direction.

Defined in this way, quantum Riemannian spaces enjoy some functorial
properties, well-known at the level of CFTs (e.g. one can take a tensor product of quantum
Riemannian spaces).

\subsection{Spectral triples, Bakry calculus and Wasserstein spaces}\label{spectral triples and Bakry calculus}

The idea  that QFTs should be studied by methods of non-commutative
geometry was suggested by Alain Connes (see [Co1]).
The idea to use Connes's approach for the description of CFTs and their degenerations
goes back to [FG].
It  was further developed in [RW] in
an attempt to interpet the earlier approach of [KoSo1] from the point of view
of Connes's spectral triples. The measure was not included in the list of data
neither in [FG] nor in [RW], since in the framework of spectral triples the measure
can be recovered from the rest of the data. Ricci curvature
was not defined in the framework of spectral triples.
In particular, it was not clear how to define a spectral triple
with non-negative Ricci curvature. Present paper can be thought of as a step in this direction.

\

Recall (see e.g. [Co1]) that a spectral triple is given by a unital $C^{\ast}$-algebra of bounded
operators in a Hilbert space and a $1$-parameter semigroup continuously acting on the space.
It is assumed that the semigroup has an infinitesimal generator $D$ which  is a positive unbounded self-adjoint operator
with the compact resolvent, and the commutator $[D,f]$ with any algebra element $f$ is bounded.

Similar structures appear in the theory
of random walks and Markov semigroups on singular spaces (see e.g. [Ba], [BaEm],
[LV], [Led], [St]). In that case one also has a probability measure which is invariant
with respect to the semigroup. This similarity makes plausible the idea that the
``abstract calculus" of Markov semigroups
developed by Bakry and \'Emery (see [BaEm], [Ba]) can be used for the description of the
topological space obtained from a collapsing
family of unitary CFTs. 

This idea was proposed by Kontsevich in a series of talks
in 2003. In those talks Kontsevich introduced the notion of  ``singular Calabi-Yau manifold" defined
in terms
of what he called Graph Field Theory (and what we call commutative Riemannian $1$-geometry below).

One hopes that
the ``moduli space of singular Calabi-Yau manifolds" with bounded dimension and fixed diameter, being equipped with a (version of) Gromov-Hausdorff
(or measured Gromov-Hausdorff)
topology, is compact. 

More generally one can expect a similar result for ``quantum Riemannian
$1$-spaces" which have bounded dimension, diameter bounded from above and Ricci curvature
bounded from below.
\footnote{ In [En]  the precompactness
of the moduli space of commutative measured Riemannian $1$-spaces  with the usual bounds on the diameter
and Ricci curvature was proved. Methods of [En] are based on explicit estimates of the heat kernel
as well as classical results by Cheeger and Colding [ChC3]. We do not see how to generalize this approach to
non-commutative case.}
To our knowledge there is no
precompactness theorem for the  space of ``abstract Bakry-\'Emery data".\footnote{Bakry-\'Emery data
naturally lead to the metric on the space of states of a $C^{\ast}$-algebra coinciding with the metric introduced by Connes's
and generalized later by Rieffel (see [Rie]) in his notion of
``quantum metric space". Although the precompactness theorem was formulated and proved by Rieffel for compact quantum metric spaces, it is not clear how to extend his approach to the case of quantum Riemannian
$1$-spaces discussed in this paper.}

In a similar vein we mention  precompactness theorems for a class of metric-measure spaces
which generalizes the class of Riemannian manifolds with non-negative Ricci curvature
(see [LV], [St]). The authors
introduced in [LV], [St] the notion of $N$-Ricci curvature, $N\in [1,\infty]$.
The notion of $N$-Ricci curvature is defined in terms
of geodesics in the space of probability measures equipped with the Wasserstein $L_2$-metric
(see Section 6).

{\it We hope that there is a non-commutative
generalization of the notion of $N$-Ricci curvature  as well as of the Wasserstein metric, so that
the class of quantum metric-measure spaces with non-negative $N$-Ricci curvature
and bounded diameter
is compact with respect to the non-commutative
generalization of the measured Gromov-Hausdorff topology
or non-commutative generalization of the metric introduced in [St].}

\

Let us make few additional remarks about the relationship of our approach with the one of Connes ( see [Co1]). In the notion of spectral triple
Connes axiomatized the triple $(A,H,\Delta)$ where $A$ is the algebra
of smooth functions on a compact closed Riemannian manifold $M$ (considered as a complex algebra with an anti-linear involution), $H=L_2(M,vol_M)$ is
the Hilbert space of functions, which are square-integrable with
respect to the volume form associated with the Riemannian metric,
and $\Delta$ is the Laplace operator associated with the metric.
\footnote{In fact Connes considered the case of spin manifolds,
so he used the Dirac operator $D$ instead of $\Delta=D^2$, and $H$ was
the space of square-integrable sections of the spinor bundle.}

From a slightly different perspective, the data are: involutive
algebra $A$, a positive linear functional $\tau(f)=\int_Mf\,vol_M$ which defines the completion
$H$ of $A$ with respect to the scalar product $\tau(fg^{\ast})$, the $\ast$-representation $A\to End(H)$,
and the 1-parameter semigroup $exp(-t\Delta), t\ge 0$ acting on $H$ by means of trace-class operators. The generator of the semigroup is a non-negative self-adjoint
unbounded operator $\Delta$ with discrete spectrum, and the algebra $A$ being naturally embedded to $H$ belongs to the domain of $\Delta$.
Thus $A$ encodes the topology of $M$, while $\tau$ encodes the measure,
and $\Delta$ encodes the Riemannian structure. Let $B_1(f,g)$ be a bilinear form $A\otimes A\to A$
given by $2B_1(f,g)=\Delta(fg)-f\Delta(g)-\Delta(f)g$.
The formula $d(\phi,\psi)=sup_{B_1(f,f)\le 1}|\varphi(f)-\psi(f)|$ defines the distance function on the space
of states of the $C^{\ast}$-completion of $A$ in terms of the spectral triple data
(the $C^{\ast}$-completion can be spelled out intrinsically in terms of the operator norm derived from the
$\ast$-representation $A\to End(H)$). Every point $x\in M$ gives rise to a state (delta-function $\delta_x$).

One  sees that the
above formula recovers the Riemannian distance function on $M$ without use of the language of points, so it can be generalized to the case of non-commutative algebra $A$.
There are many non-trivial examples of spectral triples
which do not correspond to commutative Riemannian manifolds
(see e.g. [Co1], [CoMar]). 

Let us observe that in the case of Riemannian manifolds
the 1-parameter semigroup $exp(-t\Delta), t\ge 0$ assigns a trace-class
operator $exp(-l\Delta)$ to every segment $[0,l]$, which we can view
as a very simple metrized graph with one input and one output. Moreover, the multiplication $m_A: A\otimes A\to A$  gives rise to
the family of operators $S_{l_1,l_2,l_3}$
$H^{\otimes 2}\to H$ such that

$$x_1\otimes x_2\mapsto exp(-l_3\Delta)(m_A(exp(-l_1\Delta)(x_1)\otimes exp(-l_2\Delta)(x_2))),$$
for any $l_1,l_2,l_3>0$.

In a bit more symmetric way, one has a family of trilinear
forms $H^{\otimes 3}\to \C$ such that
$(x_1,x_2,x_3)\mapsto \tau(m_A(m_A\otimes id)(exp(-l_1\Delta)(x_1)\otimes exp(-l_2\Delta)(x_2)\otimes exp(-l_3\Delta)(x_3)))$. Hence, starting with a commutative spectral triple,
we can produce trace-class operators associated with
two types of metrized graphs:

a) to a segment $I_l:=[0,l]$ we associate an operator
$exp(-l\Delta):=S_{I_l}:H\to H$, assuming that for $l=0$ we have the identity operator;

b) to the $Y$-shape graph $\Gamma_{l_1,l_2,l_3}$ with different positive lengths of the three edges we
associate an operator $S_{l_1,l_2,l_3}:=S({\Gamma_{l_1,l_2,l_3}}): H^{\otimes 2}\to H$.
Notice that $m_A(f_1\otimes f_2)=lim_{l_1+l_2+l_3\to 0}S({\Gamma_{l_1,l_2,l_3}})(f_1\otimes f_2), f_i\in A, i=1,2$,
hence the multiplication on $A$ can be recovered from operators associated to metrized graphs as long
as we assume continuity of the operators with respect to the length of an edge of the tree.

\begin{rmk}\label{remark about general graphs}
From the point
of view of Quantum Field Theory it is natural to consider more general graphs. This leads to the notion of quantum Riemannian $1$-geometry (see below).
It turns out that
this language is suitable for spelling out various differential-geometric
properties of Riemannian manifolds (non-commutative and singular in general), in particular,
the property to have the non-negative Ricci curvature.
\end{rmk}
\subsection{Possible applications}\label{applications}

One can speculate about possible applications of the ideas of this paper. The Gromov-Hausdorff
(or  measured Gromov-Hausdorff) topology is coarser
than topologies typically used in physics. 

The question is: {\it are these ``Gromov-Hausdorff type" topologies
``physical enough" to derive interesting properties of the moduli spaces?}

 For example, it is interesting
whether measured Gromov-Hausdorff topology can be useful in the study of the
so-called ``string landscape" and the problem
of finiteness of the volume of the corresponding moduli spaces of  QFTs (see e.g. [Dou 1], [Dou 2], [Dou L], [Va], [OVa]).
For example, the problem of statistics of the string vacua leads
to the counting of the number of critical points of a certain function (prepotential)
on the moduli space of certain CFTs (see e.g. [Dou 1], [Z]).
Finiteness of the volume of the moduli space
of CFTs (the volume gives the first term of the asymptotic expansion of the number of critical
points) is crucial.
According to the previous discussion, the ``Gromov-Hausdorff type" moduli space of unitary CFTs with the bounded central charge and bounded from below spectral gap
is expected to be compact. Probably the requirement that the spectrum is discrete leads to the finiteness of the
volume of the moduli space with respect to the measure derived from Zamolodchikov metric (cf. [Va]).
This would give a bound for the number of string vacua. I should also point out an interesting paper [Dou 3] which contains a discussion of the moduli of QFTs from a perspective close to  [KoSo1] and [So1].\footnote{ More recent example related to our original idea of collapsing CFT\rq{}s can be found in [OW].}

\vspace{3mm}

{\it Acknowledgements.} 
I thank to Jean-Michel Bismut, Kevin Costello, Alain Connes, Michael Douglas, Boris Feigin, Misha Gromov, Kentaro Hori, David Kazhdan, Andrey Losev, John Lott, Yuri Manin, Matilde Marcolli, Nikolay Reshetikhin, Dmitry Shklyarov, Katrin Wendland for useful conversations and correspondence. \footnote{I also thank to Hiroshi Ooguri for attracting my attention to [OW] which eventually led me to  the idea of submission of this paper  to arXiv.}
I am especially grateful to Maxim Kontsevich for numerous discussions about CFTs and  Calabi-Yau manifolds, which influenced this work very much. 
I thank to IHES for hospitality and excellent research and living conditions.
This work was partially supported by an NSF grant.

\section{Reminder on degenerating Conformal Field Theories}\label{reminder on degenerating CFTs}

This section contains the material borrowed from [KoSo1]. I include it here for motivational purposes.
\subsection{Unitary CFTs}\label{unitary CFTs}

Unitary Conformal Field
Theory is well-defined mathematically
thanks to Segal's axiomatic approach (see [Seg]).
We are going to recall Segal's axioms later.
In the case of the complex line ${\C}$ the data defining a unitary CFT
can be summarized such as follows:

1) A real number $c\ge 0$ called central charge.

2) A bi-graded pre-Hilbert {\it space of states}
$H=\oplus_{p,q\in {\bf R}_{\ge 0}}H^{p,q},p-q\in {\bf Z}$ such that
$dim(\oplus_{p+q\le E} H^{p,q})$ is
finite for every $E\in {\bf R}_{\ge 0}$. Equivalently, there is an
action of the Lie group ${\C}^{\ast}$ on $H$, so that
$z\in{\C}^{\ast} $ acts on $H^{p,q}$
as $z^p\bar{z}^q:=(z\bar{z})^p\bar{z}^{q-p}$.

3) An action of the product of Virasoro  and anti-Virasoro Lie algebras
$Vir\times \overline{Vir}$ (with the same central charge $c$)
on $H$, so that the space $H^{p,q}$
is an eigenspace for
the generator $L_0$ (resp. $\overline{L}_0$) with the eigenvalue $p$
(resp. $q$).

4) The space $H$ carries some additional
structures derived from the operator product expansion (OPE).
The OPE is described by a
linear map $H\otimes H\to H\widehat{\otimes}{\C}\{ {z,\bar{z}}\}$.
Here ${\C}\{ {z,\bar{z}}\}$ is the topological ring of formal
power series $f=\sum_{p,q}c_{p,q}z^{p}\bar{z}^{q}$ where $ c_{p,q}\in {\C},
\,p,q\to +\infty,\,\, p,q \in {\R},\, p-q\in {\Z}$.
The OPE satisfies some axioms which do not recall here
(see e.g. [Gaw]). One of the axioms is a sort of associativity of
the OPE.

Let $\phi\in H^{p,q}$. Then the number $p+q$ is called the
{\it conformal dimension} of $\phi$
(or the {\it energy}), and $p-q$ is called the {\it spin}
of $\phi$. Notice that, since the spin of $\phi$
is an integer number, the condition $p+q<1$ implies $p=q$.

The central charge $c$ can be described intrinsically by the formula
$$dim(\oplus_{p+q\le E}H^{p,q})=exp(\sqrt{4/3\pi^2cE(1+o(1))}$$
 as
$E\to +\infty$. It is expected that all possible central charges
form a countable well-ordered subset of ${\bf Q}_{\ge 0}\subset {\R}_{\ge 0}$.
If $H^{0,0}$ is a one-dimensional vector space, the corresponding CFT is called
irreducible. A general CFT
is a sum of irreducible ones. The {\it trivial} CFT has $H=H^{0,0}=\C$
and it is the unique
irreducible unitary CFT
with $c=0$.

\begin{rmk} \label{SCFT version}There is a version
of the above data and axioms for Superconformal Field Theory
(SCFT). In that case each
$H^{p,q}$ is a hermitian super vector space. There is an action
of the super extension of the product
of Virasoro and anti-Virasoro algebra on $H$.
In the discussion of the moduli
spaces below we will speak about CFTs, not SCFTs. Segal's axiomatics for SCFT
is not available from the published literature.
\end{rmk}

\subsection{Moduli space of Conformal Field Theories}\label{moduli of CFTs}

For a given CFT one can consider its group of symmetries
(i.e. automorphisms of the space $H=\oplus_{p,q} H^{p,q}$
preserving all the structures).

\begin{conj}\label{conjecture about symmetry group}
The group of symmetries is a compact Lie
group of dimension less or equal than $dim\,H^{1,0}$.
\end{conj}

Let us fix $c_0\ge 0$ and $E_{min}>0$, and consider the moduli space
${\cal M}_{c\le c_0}^{E_{min}}$ of
all irreducible CFTs with the central charge $c\le c_0$ and
$$min\{p+q>0| H^{p,q}\ne 0\}\ge E_{min}.$$

\begin{conj}\label{conjecture about dimension of miniversal deformation}
${\cal M}_{c\le c_0}^{E_{min}}$
is a compact real analytic stack  of finite local dimension.
The dimension of the base of the minimal
versal deformation of a given CFT is less or equal than $dim\,H^{1,1}$.

\end{conj}

We define
${\cal M}_{c\le c_0}=\cup_{E_{min}>0}{\cal M}_{c\le c_0}^{E_{min}}$.
We would like to compactify this stack by adding boundary components
corresponding to certain degenerations
of the theories as $E_{min}\to 0$.
The compactified space
is expected to be a compact stack $\overline{{\cal M}}_{c\le c_0}$.
In what follows we will loosely use the word ``moduli space'' instead of
the word ``stack''.

\begin{rmk}\label{Segal axioms and sigma-models}
There are basically two classes of rigorously defined CFTs: the rational
theories (RCFT) and the
lattice CFTs. They are defined algebraically (e.g. in terms
of braided monoidal categories or vertex algebras).
Physicists often consider so-called
sigma models defined in terms of maps of two-dimensional Riemannian
surfaces (world-sheets) to a Riemannian manifold (world-volume, or
target space). Such a theory depends on a choice of the Lagrangian,
which is a functional on the space of such maps.
Descriptions of sigma models as path integrals
does not have precise mathematical meaning. Segal's axioms
arose from an attempt to treat the path integral categorically.
As we will explain below, there is an alternative way to speak about
sigma models
in terms of degenerations of CFTs. Roughly speaking, sigma models
``live" near the boundary of the compactified moduli space $\overline{{\cal M}}_{c\le c_0}$.

\end{rmk}

\subsection{Physical picture of a simple collapse}\label{physics of simple collapse}

In order to compactify ${\cal M}_{c\le c_0}$
we consider degenerations of CFTs as $E_{min}\to 0$.
A degeneration is given by a one-parameter (discrete or continuous)
family $H_{\varepsilon}, \varepsilon\to 0$
of bi-graded spaces as above,
where $(p,q)=(p(\varepsilon),q(\varepsilon))$,
equipped with OPEs, and  such that $E_{min}\to 0$.
The subspace of fields with
conformal dimensions  vanishing as $\varepsilon\to 0$
gives rise to a commutative algebra $H^{small}=\oplus_{p(\varepsilon)\ll 1}
H_{\varepsilon}^{p(\varepsilon),p(\varepsilon)}$
(the algebra structure is given by the leading terms
in OPEs).
The spectrum $X$ of $H^{small}$ is expected to be a compact space
(``manifold with
singularities'')  such that $dim\,X\le c_0$.
It follows from the conformal invariance and the OPE, that
the grading of $H^{small}$ (rescaled as $\varepsilon\to 0$)
is given by the eigenvalues of a
second order differential operator defined on the smooth part of $X$.
The operator has positive eigenvalues and is determined
up to multiplication by a scalar. This implies that the smooth part of $X$
carries a metric $g_X$, which is
also defined up to multiplication by a scalar. Other terms in OPEs give rise
to additional differential-geometric structures on $X$.

 Thus, as a first approximation to the real picture, we assume
 the following description of a ``simple collapse'' of a family of CFTs.
The degeneration of the family is described by the point of the boundary of
$\overline{{\cal M}}_{c\le c_0}$ which is a
triple $(X, {\R}_+^{\ast}\cdot g_X,\phi_X)$, where the
 metric $g_X$  is defined up to
a positive scalar factor, and $\phi_X:X\to {\cal M}_{c\le c_0-dim\,X}$
is a map. One can have some extra conditions on the data.
For example, the metric $g_X$ can satisfy the Einstein equation.

Although the scalar factor for the metric is arbitrary, one should
imagine that the curvature of $g_X$ is ``small'', and the injectivity
radius of $g_X$ is ``large''.
The map $\phi_X$ appears naturally from the point
of view of the simple collapse of CFTs described above.
Indeed, in the limit $\varepsilon\to 0$, the space $H_{\varepsilon}$
becomes an $H^{small}$-module. It can be thought of as a space
of sections of an infinite-dimensional vector bundle
$W\to X$. One can argue that fibers of $W$ generically are
spaces of states of CFTs with central charges less or equal
than $c_0-dim\,X$.
 This is encoded in the map $\phi_X$. In the case when CFTs
from $\phi_X(X)$ have non-trivial symmetry groups, one expects
a kind of a gauge theory on $X$ as well.

Purely bosonic sigma-models correspond to the case when
$c_0=c(\varepsilon)=dim\,X$
and the residual theories (CFTs in the image of $\phi_X$) are all trivial.
The target space $X$ in this case should carry a Ricci flat metric.
In the $N=2$ supersymmetric case the target space $X$ is a Calabi-Yau manifold,
and the residual bundle of CFTs is a bundle of free fermion theories.

\begin{rmk}\label{targets for sigma-models}
It was conjectured in [KoSo1] that {\it all compact Ricci flat manifolds
(with the metric defined up to a constant scalar factor) appear as target
spaces of degenerating CFTs}. Thus, the construction of the compactification
 of the moduli space of CFTs should include a compactification
 of the moduli space of Einstein manifolds.
 As we already mentioned in the Introduction, there is a deep relationship between the
 compactification of the moduli space of CFTs
 and  Gromov's compactification.
Moreover, as we will discuss below,
all target spaces appearing as
limits of CFTs have in some sense non-negative Ricci curvature. More precisely,
the limit of the rescaled Virasoro operator $L_0+\overline{L}_0$ satisfies Bakry curvature-dimension
inequality $CD(0,\infty)$. In the case of compact Riemannian manifolds the latter is equivalent
to non-negativeness of Ricci curvature.
\end{rmk}

\subsection{Multiple collapse and the structure of the boundary}\label{multiple collapse}

In terms of the Virasoro operator $L_0+\overline{L}_0$ the degeneration of CFT
(`` collapse") is described by a subset (cluster) $S_1$
in the set of eigenvalues of $L_0+\overline{L}_0$
which approach to zero ``with the same speed'' provided $E_{min}\to 0$.
The next level of the collapse is described by
another subset $S_2$ of eigenvalues
of $L_0+\overline{L}_0$. Elements of $S_2$ approach to zero ``modulo the first collapse"
(i.e. at the same speed, but ``much slower'' than elements of $S_1$).
One can continue to build a tower of degenerations.
It leads to an hierarchy of
boundary strata. Namely, if there are further degenerations
of CFTs parametrized by $X$, one gets a fiber bundle
over the space of triples
$(X,{\R}_+^{\ast}\cdot g_X,\phi_X)$ with the
fiber which is the space of triples of similar sort.
Finally, we obtain the following
qualitative geometric picture of the boundary
$\partial \overline {{\cal M}}_{c\le c_0}$.

A boundary point is given by the following data:

1) A finite tower of maps of compact topological spaces
$p_i:\overline{X}_i\to
\overline{X}_{i-1}, 0\le i\le k$, $\overline{X}_0=\{pt\}$.

2) A sequence of smooth manifolds
$({X}_i,g_{{X}_i}),
0\le i\le k$, such that ${X}_i$ is a dense
subspace of $\overline{X}_i$, and $dim\,X_i>dim\,X_{i-1}$,
and $p_i$ defines a fiber bundle $p_i:X_i\to X_{i-1}$.

3) Riemannian metrics on the fibers of the restrictions
of $p_i$ to $X_i$, such that the diameter of each fiber is finite.
In particular the diameter of $X_1$ is finite, because it is the
only fiber of the map $p_1:X_1\to \{pt\}$.

4) A map $X_k\to {\cal M}_{c\le c_0-dim\,X_k}$.

The data above are considered up to the natural action
of the group $({\R}_+^{\ast})^{k}$ (it rescales the metrics on fibers).

There are some additional data, like non-linear connections
on the bundles $p_i:X_i\to X_{i-1}$. The set of data should
satisfy some conditions, like differential equations on the metrics.
It is an open problem to describe these conditions  in general case.
In the case of $N=2$ SCFTs corresponding to
sigma models with Calabi-Yau target spaces these geometric conditions
were formulated as a conjecture in [KoSo1].

\subsection{Example: Toroidal models }\label{toroidal models}

Non-supersymmetric toroidal model is described by the
so-called Narain lattice, endowed with some additional data.
More precisely, let us fix the central charge $c=n$
which is a positive integer number. What physicists call the Narain lattice $\Gamma^{n,n}$
is a unique unimodular lattice of
rank $2n$ and the signature $(n,n)$. It can be described
as ${\Z}^{2n}$ equipped with the quadratic form
$Q(x_1,...,x_n,y_1,...,y_n)=\sum_ix_iy_i$.
The moduli space of toroidal CFTs is given by

$${\cal M}_{c=n}^{tor}=O(n,n,{\Z})\backslash O(n,n,{\R})/O(n,{\R})
\times O(n,{\R}).$$

Equivalently, it is a quotient  of the open part of
the Grassmannian
$\{V_+\subset {\R}^{n,n}|\, dim\,V_+=n,Q_{|V}>0\}$
by the action of $O(n,n,{\Z})=Aut(\Gamma^{n,n},Q)$.
Let $V_-$ be the orthogonal complement to $V_+$.
Then every vector of $\Gamma^{n,n}$
can be uniquely written as $\gamma=\gamma_{+}+\gamma_{-}$,   where
 $\gamma_{\pm}\in V_{\pm}$.
For the corresponding CFT one has
$$\sum_{p,q}dim(\,H^{p,q})z^p\bar{z}^q=
\bigl|\prod_{k\ge 1}(1-z^k)\bigr|^{-2n}\sum_{\gamma\in\Gamma^{n,n}}
z^{Q(\gamma_{+})}\bar{z}^{\,-Q(\gamma_{-})}$$

Let us describe the (partial) compactification of the moduli space ${\cal M}_{c=n}^{tor}$
by collapsing toroidal CFTs.  Suppose that we have a   one-parameter
family of toroidal theories such that  $E_{min}(\varepsilon)$ approaches zero. Then for corresponding
vectors in $H_{\varepsilon}$  one gets
$p(\varepsilon)=q(\varepsilon)\to 0$. It implies that
$Q(\gamma(\varepsilon))=0, Q(\gamma_{+}(\varepsilon))\ll 1$.
It is easy to see that
the vectors $\gamma(\varepsilon)$
form a semigroup with respect to addition. Thus one obtains
a (part of) lattice of the rank less or equal than $n$.
In the case of ``maximal'' simple collapse the rank will be equal to $n$.
One can see that the corresponding points of the boundary give
rise to the following data: $(X,{\R}_+^{\ast}\cdot g_X, \phi_X^{triv};B)$, where
$(X,g_X)$ is a flat $n$-dimensional torus, $B\in H^2(X,i{\R}/{\Z})$
and $\phi_X^{triv}$ is the constant map form $X$ to the trivial theory point in the moduli space of CFTs.
These data in turn give rise to a toroidal CFT, which can be realized
as a sigma model with the target space $(X,g_X)$ and
given B-field $B$. The residual bundle of CFTs on $X$ is trivial.

Let us consider a $1$-parameter family of CFTs given by
$(X,\lambda g_X, \phi_X^{triv};B=0)$,
where $\lambda \in (0,+\infty)$.
There are two degenerations of this family, which define
two points of the boundary $\partial \overline{\cal M}_{c=n}^{tor}$.
 As $\lambda \to +\infty$,
we get a toroidal CFT defined by $(X,{\R}_+^{\ast}\cdot g_X, \phi_X^{triv};B=0)$.
As $\lambda \to 0$ we get $(X^{\vee},{\R}_+^{\ast}\cdot g_{X^{\vee}}, \phi_X^{triv};B=0)$,
where $(X^{\vee},g_{X^{\vee}})$ is the dual flat torus.

There might be further degenerations of the lattice.
Thus one obtains a stratification of the compactified moduli space
of lattices (and hence  CFTs).
Points of the compactification are described by flags of vector spaces
$0=V_0\subset V_1\subset V_2\subset...\subset V_k\subset {\R}^n$.
In addition one has a lattice $\Gamma_{i+1}\subset V_{i+1}/V_i$,
considered up to a scalar factor. These data give rise to
a tower of torus bundles $X_k\to X_{k-1}\to...\to X_1\to \{pt\}$
over tori with fibers $(V_{i+1}/V_i)/\Gamma_{i+1}$.
If $V_k\simeq {\R}^{n-l}, l\ge 1$, then one has also
a map from the total space $X_k$
of the last torus bundle to the point $[H_k]$ in the moduli space
of toroidal theories of smaller central charge:
 $\phi_n:X_k\to {\cal M}_{c=l}^{tor}$,
$\phi_k(X_k)=[H_k]$.

\subsection{Example: WZW model for $SU(2)$}\label{WZW model}

In this case we have a discrete family with
 $c={3k\over{k+2}}$, where $k\ge 1$ is an integer number
called {\it level}.
In the limit $k\to +\infty$ one gets $X=SU(2)=S^3$
equipped with the standard metric.
The corresponding bundle is the trivial bundle of
trivial CFTs (with $c=0$ and
$H=H^{0,0}={\C}$). Analogous picture holds for an arbitrary
compact simply connected simple group $G$.

\subsection{Example: minimal models}\label{minimal models}

This example has been worked out in [RW], Section 4.
One has a sequence of unitary CFTs $H_m$ with the central charge
$c_m=1-{6\over{m(m+1)}}, m\to +\infty$. In this case $c_m\to 1$,
and the limiting space is the interval $[0,\pi]$.
The metric is
given by $g(x)={4\over \pi^2}sin^4x, x\in [0,\pi]$.
The corresponding volume form is  $vol_g=\sqrt{g(x)}dx={2\over \pi}sin^2xdx$.
In all above examples the volume form on the
limiting space is the one associated with the Riemannian metric.

\subsection{A-model and B-model of $N=2$ SCFT as boundary strata}\label{A and B model}

In the case of $N=2$ Superconformal Field Theory one can modify the above
considerations in the natural way. As a result one arrives to the following
picture of the simple collapse.

The boundary of the compactified moduli space
$\overline {{\cal M}}^{N=2}$ of $N=2$ SCFTs with a given central charge
contains an open
stratum given by sigma models with Calabi-Yau targets.
Each stratum is parametrized by the classes of equivalence of
quadruples $(X,J_X,{\R}_+^{\ast}\cdot g_X,B)$ where $X$ is
a compact real manifold, $J_X$ a complex structure,
$g_X$ is a Calabi-Yau metric,
 and $B\in H^2(X,i{\bf R}/{\bf Z})$ is a
$B$-field. The residual bundle of CFTs is a bundle of free fermion
theories.

 As a consequence of supersymmetry,
the moduli
space ${\cal M}^{N=2}$  of superconformal field theories is a complex
manifold which
is locally
isomorphic to the product of two complex manifolds. \footnote{Strictly speaking,
one should exclude models with chiral fields of conformal dimension $(2,0)$,
 e.g. sigma models on hyperk\"ahler manifolds.}
It is believed that this decomposition (up to certain
corrections) is global.
Also, there are two types of sigma models with Calabi-Yau targets:
$A$-models and $B$-models.
Hence, the traditional picture
of the compactified moduli space looks as follows:

\begin{figure}[h!]
\centering
\includegraphics{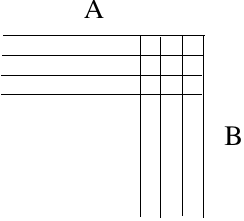}

\end{figure}

Here the boundary consists of two open strata (A-stratum and B-stratum)
and a mysterious meeting point. This point corresponds,
in general, to a submanifold
of codimension one in the closure of A-stratum and of B-stratum.

As we explained in [KoSo1] this picture should be modified. Namely, there is another open
stratum of $\partial \overline {{\cal M}}^{N=2}$ (called T-stratum in [KoSo1]).
 It consists of toroidal
 models (i.e. CFTs  associated with Narain lattices),
 parametrized by a
 manifold $Y$ with a Riemannian metric defined up to a scalar factor.
 This $T$-stratum meets both $A$ and $B$ strata along the codimension
 one stratum corresponding to the double collapse.
 Therefore the ``true'' picture
 is obtained from the traditional one
 by the real blow-up at the corner:

\begin{figure}[h!]

\centering
\includegraphics{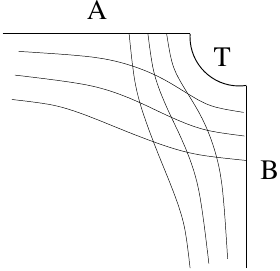}

\end{figure}
\subsection{Mirror symmetry and the collapse}\label{MS and collapse}

Mirror symmetry is related to the existence
of two different strata of the boundary $\partial \overline {{\cal M}}^{N=2}$
which we called A-stratum and B-stratum.
As a corollary, same quantities
admit different geometric descriptions near different
strata. In the traditional picture, one can introduce
natural coordinates in a small neighborhood of a boundary point
corresponding to $(X,J_X,{\R}_+^{\ast}\cdot g_X,B)$.
Skipping $X$ from the notation, one can say that the coordinates
are $(J,g,B)$ (complex structure, Calabi-Yau metric and the B-field).
Geometrically, the pairs $(g,B)$ belong to the preimage of the
K\"ahler cone under the natural map $Re: H^2(X,{\C})\to H^2(X,{\R})$
(more precisely, one should consider $B$ as an element
of $H^2(X,i{\R}/{\Z})$). It is usually said, that one considers
an open domain in the complexified K\"ahler cone with the property that it contains together
with the class of metric $[g]$  also the ray
$t[g], t\gg 1$. The mirror symmetry gives rise to an
identification of neighborhoods of $(X,J_X,{\R}_+^{\ast}\cdot g_X,B_X)$
and $(X^{\vee},J_{X^{\vee}},{\R}_+^{\ast}\cdot g_{X^{\vee}},B_{X^{\vee}})$
such that $J_X$ is interchanged with $[g_{X^{\vee}}]+iB_{X^{\vee}})$
and vice versa.

We can describe this picture in a different
way.
Using the identification of complex and K\"ahler moduli, one can choose
$([g_X],B_X, [g_{X^{\vee}}],B_{X^{\vee}})$ as local coordinates near
the meeting point of A-stratum and B-stratum.
 There is an action of the additive
semigroup ${\R}_{\ge 0}\times {\R}_{\ge 0}$
in this neighborhood. It is given explicitly by the formula
$([g_X],B_X, [g_{X^{\vee}}],B_{X^{\vee}})\mapsto
 (e^{t_1}[g_X],B_X, e^{t_2}[g_{X^{\vee}}],B_{X^{\vee}})$ where
 $(t_1,t_2)\in {\R}_{\ge 0}\times {\R}_{\ge 0}$.
 As $t_1\to +\infty$, a point of the moduli space
 approaches  the B-stratum,
 where the metric is defined up to a positive scalar only.
 The action of the second semigroup ${\R}_{\ge 0}$ extends
by continuity to the non-trivial action on the B-stratum.
Similarly, in the limit
 $t_2\to +\infty$ the flow retracts the point to the A-stratum.

This picture should be modified, if one makes a real blow-up at
the corner, as we discussed before. Again, the action of the semigroup
${\R}_{\ge 0}\times {\R}_{\ge 0}$ extends continuously to the
boundary.
Contractions to the A-stratum and B-stratum carry non-trivial
actions of the corresponding semigroups isomorphic to
${\R}_{\ge 0}$. Now, let us choose a point
in, say, A-stratum. Then the semigroup flow takes it
along the boundary to
the new stratum, corresponding to the double collapse.
The semigroup ${\R}_{\ge 0}\times {\R}_{\ge 0}$ acts trivially on this stratum.
A point of the double collapse is also
a limiting point of a $1$-dimensional orbit of
${\R}_{\ge 0}\times {\R}_{\ge 0}$ acting on the T-stratum.
Explicitly, the element $(t_1,t_2)$ changes the size
of the tori defined by the Narain lattices, rescaling them with the coefficient
$e^{t_1-t_2}$. This flow carries the point of T-stratum
 to another point
of the double collapse, which can be moved then inside of the B-stratum.
The whole path, which is the intersection of $\partial \overline{\cal M}^{N=2}$
and the ${\R}_{\ge 0}\times {\R}_{\ge 0}$-orbit,
connects an A-model with the corresponding B-model
through the stratum of toroidal models. We can depict it as
follows:

\begin{figure}[h!]

\includegraphics{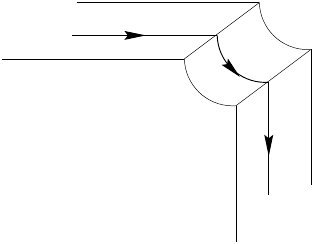}

\end{figure}

The conclusion of the above discussion is the following:
in order to explain the mirror symmetry phenomenon
it is not necessary to build full SCFTs.
It is sufficient to work with simple
toroidal models on the boundary of the compactified moduli
space $\overline {{\cal M}}^{N=2}$.

\section{Segal's axioms and collapse}\label{Segal\rq{}s axioms and collapse}

\subsection{Segal's axioms}\label{Segal\rq{}s axioms}

Let us recall Segal's axioms of a  2-dimensional unitary  CFT (see e.g. [FFRS], [Gaw], [Ru], [Seg]).
Similarly to [FFRS] and [Ru] we use surfaces with a metric rather than surfaces with complex structure.
To simplify the description we use oriented surfaces only.
The data of a unitary CFT are described such as follows:

1) Real number $c$ called {\it central charge}.

2) $2$-dimensional {\it world-sheet}. The latter is defined by:

2a) an oriented 2-dimensional manifold $\Sigma$, possibly disconnected,
with finitely many boundary components $(C_i)_{i\in I}$ labeled by elements
of the disjoint union of finite sets: $I=I_{-}\sqcup I_{+}$. Components
labeled by the elements of $I_{-}$ (resp. $I_{+}$) are called {\it incoming} (resp. {\it outcoming}).
The orientation of $\Sigma$ induces orientations of all $C_i$ as boundary components.
This orientation is called canonical;

2b) a Riemannian metric $g_{\Sigma}$ on $\Sigma$;

2c) for some $\varepsilon >0$ (which depends on $i\in I$) a
real-analytic conformal embedding to $(\Sigma, g_{\Sigma})$ of either the flat annulus $A_{\varepsilon}^{-}=
\{z\in \C|1-\varepsilon<|z|\le 1\}$ (for $i\in I_{-}$)  or the flat annulus $A_{\varepsilon}^{+}=
\{z\in \C|1\le |z|<1+\varepsilon\}$ (for $i\in I_{+}$)
(the annuli are equipped with the standard flat metric ${dz\,d\overline{z}\over{z\overline{z}}}$) such that:
$f_i^{\pm}(|z|=1)=C_i$ and
$f_i^{-}$ (resp $f_i^{+}$) is orientation preserving (resp. reversing) with respect to the canonical
orientation.
The map $f_i^{+}$ (resp.$f_i^{-}$) is called a {\it parametrization} of $C_i$.

3) Complex separable Hilbert space $H$ \footnote{This condition can be relaxed so one can assume that $H$
is a locally compact vector space, e.g. a nuclear vector space.} equipped with an antilinear involution
$x\mapsto \sigma({x})$.

4) A trace-class operator ({\it amplitude})
$S(\Sigma,(f_i)_{i\in I},g_{\Sigma}):\otimes_{i\in I_{-}}H\to \otimes_{i\in I_{+}}H$
(by convention the empty tensor product is equal to $\C$). We will sometimes denote by $H_i$
the tensor factor corresponding to $i\in I$.

These data are required to satisfy the following axioms:

CFT 1) If $(\Sigma,(f_i),g_{\Sigma})=\sqcup_{\alpha} (\Sigma_{\alpha},(f_i^{\alpha}),g_{\Sigma_{\alpha}})$ then
$S(\Sigma,(f_i),g_{\Sigma})=
\otimes_{\alpha}S(\Sigma_{\alpha},(f_i^{\alpha}),g_{\Sigma_{\alpha}})$.

CFT 2) Let $i_0\in I$, and $\overline{f}_{i_0}(z,\overline{z})=f_{i_0}({1\over{z}},{1\over{\overline{z}}})$
(i.e. $\overline{f}_{i_0}$ induces the opposite orientation on $C_{i_0}$). For the new world-sheet
we require that if $i_0$ was in $I_{-}$ (resp. $I_+$) then it is now in $I_{+}$ (resp. $I_{-}$). The
condition says:
$\langle S(\Sigma,\overline{f}_{i_0},(f_i)_{i\in I\setminus i_0},g_{\Sigma})x_{i_0}\otimes x,y\rangle=\langle S(\Sigma,(f_i)_{i\in
I},g_{\Sigma})(\sigma({x}_{i_0})\otimes x, y\rangle,$
where $\langle,\rangle$ denotes the hermitian scalar product on $H$.

CFT 3) Let $i_0\in I_{-}, j_0\in I_{+}$, and $f_{i_0}$ and $f_{j_0}$ be the corresponding
parametrizations. Let us change (by a real-analytic change of coordinates in the annulus $A_{\varepsilon}^{+}$) the
parametrization $f_{j_0}$ in such a way that the pull-backs of the metric $g_{\Sigma}$
under $f_{i_0}$ and $f_{j_0}$ coincide at the corresponding points of the circle
$|z|=1\subset A_{\varepsilon}^{\pm}$. Let us keep the same notation
$f_{j_0}$ for the new parametrization. Identifying points $f_{i_0}(z,\overline{z})$ and $f_{j_0}(z,\overline{z})$,
for $|z|=1$ we obtain a new $2$-dimensional oriented surface
$\Sigma_{i_0,j_0}$ such that
$C_{i_0},i_0\in I_{-}$ and $C_{j_0}, j_0\in I_{+}$ are isometrically identified. By construction,
the surface  $\Sigma_{i_0,j_0}$ carries a smooth
metric induced by $g_{\Sigma}$.
In order to formulate the next condition we need to introduce the notion of partial trace. Let $A: V\otimes H_{i_0}\to W\otimes H_{j_0}$ be a linear map. Using an antilinear involution $\sigma$
let us identify $H_{j_0}=H$ with the dual space $H_{i_0}^{\ast}=H^{\ast}$ such as follows: $y\mapsto l_y= \langle \bullet, \sigma(y)\rangle$.
Then we  define
$Tr_{i_0,j_0}(A): V\to W$ by the formula $Tr_{i_0,j_0}(A)(v\otimes x_{i_0})=\sum_m l_{e_m}(A(v\otimes x_{i_0}))$,
where the sum is taken over elements $e_m$ of an orthonormal basis of $H_{j_0}$.
The condition says: $S(\Sigma_{i_0,j_0},(f_i)_{i\in I\setminus\{i_0,j_0\}},g_{\Sigma_{i_0,j_0}})=Tr_{i_0,j_0}S(\Sigma,(f_i)_{i\in I},g_{\Sigma})$.

CFT 4) Let $\overline{\Sigma}$ denotes the same $2$-dimensional manifold $\Sigma$, but with the orientation
changed to the opposite one, and with $I_{-}$ being interchanged with $I_{+}$, but all parametrizations remained
the same. Using the involution $\sigma$ we identify each dual space $H_i^{\ast}$ with $H_i, i\in I$ as above.
The condition says that the operator corresponding to $\overline{\Sigma}$ coincides
with the dual operator  $S(\Sigma,(f_i),g_{\Sigma})^{\ast}: \otimes_{i\in I_{+}}H\to \otimes_{i\in I_{-}}H$.

CFT 5) If the metric $g_{\Sigma}$ is replaced by $e^hg_{\Sigma}$ where
$h$ is a real-valued smooth function, then
$$S(\Sigma,(f_i),e^hg_{\Sigma})=exp({c\over {96 \pi}}D(h))S(\Sigma,(f_i),g_{\Sigma}),$$

where
$$D(h)=\int_{\Sigma}(|\nabla\,h|^2+4Rh)d\mu_{\Sigma},$$
$R$ is the scalar curvature of the metric $g_{\Sigma}$, $d\mu_{\Sigma}$
is the measure corresponding to $g_{\Sigma}$, and $\nabla\,h$ is the gradient of $h$
with respect to  $g_{\Sigma}$.

CFT 6) The operator $S(\Sigma,(f_i),g_{\Sigma})$ is invariant with respect to isometries of world-sheets
which respect  labelings and parametrizations of the boundary components.

\subsection{Collapse of CFTs as a double-scaling limit}\label{double-scaling limit}

Collapse of a family of unitary CFTs admits a description in the language of Segal's axioms.
Let us consider the set ${\cal W}$ of isomorphism classes of wordlsheets defined in the previous subsection.
An isomorphism of wordlsheets is an isometry of two-dimensional manifolds with  boundary, which respects
separate labelings of incoming and outcoming circles as well as parametrizations.

For a fixed a non-negative integer number $g\ge 0$ let us consider a subset ${\cal P}_g\subset {\cal W}$
consisting of worldsheets $\Sigma$ which are  surfaces of genus $g\ge 0$
glued from a collection of spheres with three holes (a.k.a. pants) joined by flat cylinders. The number of cylinders depend
on the genus $g$ only. Every worldsheet is conformally equivalent to the one like this

\begin{figure}[h!]
\centering
%\scalebox{0.5}
{\includegraphics[scale=0.5]{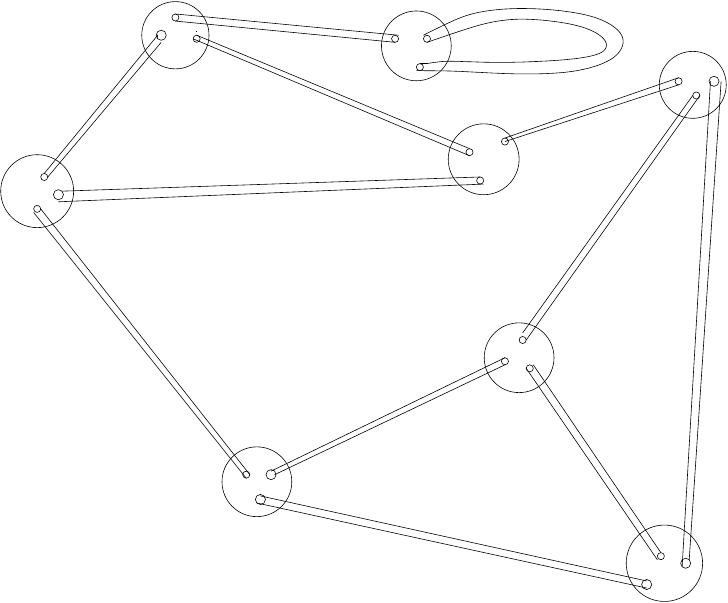}}

\end{figure}
\FloatBarrier

Without giving a definition of the topology on the moduli space
of unitary CFTs we would like to describe ``a path to infinity" in the moduli space.

Namely, let us consider the amplitudes for a family of unitary CFTs with the same central charge $c$,
depending on the parameter $\varepsilon\to 0$,
which satisfy the following properties:

1) The minimal eigenvalue $\lambda_{min}(\varepsilon)$ of the non-negative operator $L_0+\overline{L}_0$ corresponding to the standard cylinder $S^1_{R=1}\otimes [0,1]$ satisfies the property
$\lambda_{min}(\varepsilon)=\lambda_0\varepsilon\to 0+o(\varepsilon)$.

2) Consider a family of worldsheets $\Sigma=\Sigma(l_1,...,l_m,R_1,...,R_m)$ of genus $g$ such that
$min_i\,l_i\to +\infty, max_iR_i/l_i\to 0$ and which are glued from $2$-dimensional spheres with three holes each,
connected by $m$ ``long tubes", which are flat cylinders each of which is isometric to $S^1_{R_i}\times [0,l_i]$ (see Figure below), where $S^1_{R_i}$ is the circle of radius $R_i$, $1\le i\le m$.
Then  the following holds

$$l_i\lambda_{min}(\varepsilon)\to l_i^0<\infty$$
as $l_i\to\infty,\varepsilon\to 0$ uniformly for $1\le i\le m$.

Let us normalize the partition function $Z$ of the standard sphere $S^2$ in such a way that $Z(S^2)=1$.

\begin{conj}\label{bound on partition function} For the family of closed surfaces  $\Sigma$ as above we have

$$Z(\Sigma)exp(-{c\over 6}\cdot{l_i\lambda_{min}})\le const,$$

where the constant does not depend on $\Sigma$.

\end{conj}

In  the next conjecture we use the terminology
of subsequent sections.

\begin{conj} \label{conjecture about the limit}After rescaling $L_0+\overline{L}_0$ by the factor
$\lambda_{min}^{-1}$, there is a limit in the sense of Section \ref{spaces with measure} of
the unitary CFTs, which is a quantum Riemannian $1$-space in the sense
of Section \ref{quantum Riemannian 1-spaces}, such that the  operator $L$ obtained as a limit of $(L_0+\overline{L}_0)/\lambda_{min}$  satisfies
the curvature-dimension inequality $CD(0,\infty)$ from Section \ref{semigroups and CD inequalities}.
\end{conj}

In this sense CFTs collapse to QFTs with the space-time having non-negative Ricci curvature.

\section{Quantum Riemannian $d$-geometry}\label{quantum Riemannian geometry}

\subsection{$2$-dimensional case and CFT}\label{2d case}

Here we follow [Ru]. 

For any $\varepsilon>0$ we denote by
$A_{\varepsilon}$ the annulus $\{z\in \C|1-\varepsilon<|z|<1+\varepsilon\}=A_{\varepsilon}^{-}\cup A_{\varepsilon}^{+}$.
Let consider the (non-unital) category $Riem_2$ whose objects are $(k+1)$-tuples $X=(\varepsilon; f_1,...,f_k), k\ge 0$,
where $\varepsilon>0$ and $f_i:A_{\varepsilon}\to \R_{>0}, 1\le i\le k$ is a smooth function.

Notice that each function $f_i$ defines a metric $f_i(x,y)(dx^2+dy^2)$ on $A_{\varepsilon}$.
Let $Y=(\eta; g_1,...,g_l)$ be another object of $Riem_2$. A {\it morphism} $\phi:X\to Y$ is by definition
a triple $(M,g_{M}, j_{-},j_{+})$, where $(M,g_M)$ is a $2$-dimensional compact oriented
Riemannian manifold with the boundary, and $j_{-}:\sqcup_{1\le i\le k}A_{\varepsilon}^{-}\to M$ and
$j_{+}:\sqcup_{1\le i\le l}A_{\eta}^{-}\to M$ orientation
preserving (reversing for $j_{+}$) isometric embeddings  such that $Im(j_{-})\cap Im(j_{+})=\emptyset$,
and $\partial M$ is the union of the images of $S^1=\{z\in A_{\varepsilon}||z|=1\}$ under $j_{\pm}$.
Embeddings $j_{\pm}$ are called {\it parametrizations}. Boundary components of $M$ parametrized by $j_{-}$
(resp. $j_{+}$) are called {\it incoming} (resp. {\it outcoming}). 

Composition of morphisms $\phi\circ \psi$
is defined in the natural way: one identifies the point $j_{+}(z), z\in S^1$ on the outcoming boundary component of the surface defined by $\psi$ with the point $j_{-}(z), z\in S^1$ on the corresponding incoming boundary component of the surface defined by $\phi$. 

One makes $Riem_2$ into a rigid symmetric monoidal category via disjoint union
operation on objects: $(\varepsilon_1, f_1)\otimes (\varepsilon_2, f_2)=
(\varepsilon, f_1,f_2)$, where $\varepsilon=min \{\varepsilon_1,\varepsilon_2\}$ .
The duality functor is given by $(\varepsilon, f(z))^{\ast}= (\varepsilon, \overline{f(\overline{z})})$,
where the bar means complex conjugation.  On morphisms the tensor product is given by the disjoint union of surfaces together with taking the
minimal $\varepsilon$ for all incoming and outcoming boundary components. Finally, for a morphism
$\phi: X\otimes Y\to Y\otimes Z$ one has a trace map $tr_Y(\phi): X\to Z$ obtained by identification
of points $j_{+}(z), z\in S^1$ and $j_{-}(z), z\in S^1$ of outcoming and incoming boundary components
of $Y$.

\begin{defn} \label{nc Riemannian 2-space} A (commutative) Riemannian $2$-space is a monoidal functor $F: Riem_2\to Hilb_{\C}$
to the rigid monoidal category of separable complex Hilbert spaces

\end{defn}
As usual, the tensor product of Hilbert spaces is defined as the completed
algebraic tensor product. We will denote it simply by $\otimes$, changing sometimes the notation
to $\widehat{\otimes}$ in order to avoid a confusion.

A commutative Riemannian $2$-space defines a unitary CFT with the central charge $c\in \R$ if $F$ is  a functor to the category
of complex Hilbert spaces satisfying the following properties:

a) $F$ depends on the isometry class of a surface defined by a morphism, not by the surface itself. It assigns the same
Hilbert space $H$ to all objects $(\varepsilon;f_1,...,f_k)$;

b) $F$ preserves trace maps;

c) if $\phi:X\to Y$ is a morphism and $\overline{\phi}:Y\to X$ is obtained by
reversing the orientation of the corresponding surface then $F(\overline{\phi})$
is the Hermitian conjugate to $F(\phi)$;

d) change of the metric $g_M\mapsto e^hg_M$ on the surface (recall that $h$ is a
smooth real function) defined by a morphism
$\phi=(M,g_{M}, j_{-},j_{+})$ leads to a new morphism $\phi_h$ (between the same objects
according to a)) such that $F(\phi_h)=exp({c\over {96 \pi}}D(h))F(\phi)$, and
$D(h)$ is the Liouville action described in CFT 5) in the previous Section.

In order to define a {\it quantum Riemannian $2$-space}, we start with the category
$Riem_2^{NC}$ with the  objects which are $(k+1)$-tuples $(\varepsilon;f_1,...,f_k)$, as before.
A morphism is defined as a quadruple $(M,g_M,j_+,j_-,p)$ where $(M,g_M,j_+,j_-)$ is
a Riemannian $2$-dimensional surface with parametrized neighborhood of the boundary, as before,
and $p$ is a {\it marking} of the incoming (resp. outcoming) boundary components of $\partial M$
by finite sets $\{1,...,k\}$ (resp. $\{1,...,l\}$).
On the set of objects we now introduce a rigid {\it monoidal} structure, which is not symmetric.
The tensor product $(\varepsilon_1, f_1)\otimes (\varepsilon_2, f_2)$ is defined
as $(min\{\varepsilon_1,\varepsilon_2\} ,f_1,f_2)$, as before. But now there is no
commutativity isomorphism
$X_1\otimes X_2\to X_2\otimes X_1$ if $f_1$ is not equal to $f_2$, since isomorphisms
in $Riem_2^{NC}$ must respect markings.
A quantum Riemannian $2$-space is defined by
a monoidal functor $F: Riem_2^{NC}\to Hilb_{\C}$.
In general $F$
does not commute with the natural action of the product of symmetric groups $S_k\times S_l$
on the markings by the sets $\{1,...,k\}$ (incoming circles) and $\{1,...,l\}$
(outcoming circles).

\subsection{General case and spaces with measure}\label{spaces with measure}

Let us fix a non-negative integer $d$. A {\it quantum Riemannian $d$-space}
is defined as a monoidal functor $F:Riem_d^{NC}\to Hilb_{\C}$, where $Riem_d^{NC}$
is a rigid monoidal category of marked Riemannian manifolds described below (cf. [Seg], comments to
Section 4):

a) An object of $Riem_d^{NC}$ is an  isometry class of a neighborhood of a compact oriented Riemannian
$(d-1)$-manifold in an oriented $d$-manifold (called {\it Riemannian $d$-germ}) together with
a bijection between the set of its connected components  and a finite set
$\{1,...,n\}$ for some $ n\ge 1$.

b) Morphisms are oriented compact Riemannian $d$-dimensional bordisms.
The Riemannian metric is trivialized near the boundary as the product of the given
metric on the boundary and the standard flat metric on the interval (this what we called
parametrization in the two-dimensional case).
Composition of morphisms
is defined via pasting and gluing operation which respects to the metric, similarly to the case $d=2$ considered above.

c) Tensor product $X_1\otimes X_2$ is defined as the disjoint union of Riemannian $d$-germs $X_1\sqcup X_2$
equipped with the marking such that connected components of $X_1$ are marked first.
In this way $X_1\otimes X_2$ is not necessarily isomorphic to $X_2\otimes X_1$.
Duality corresponds to the reversing of the orientation.

Similarly to the two-dimensional case we have an operation $tr_Y(\phi):X \to Z$
associated with a morphism $\phi: X\otimes Y\to Y\otimes Z$.

The Hilbert space $H_X:=F(X)$ is called the {\it space of states} assigned to $X$, for a morphism $f:X\to Y$ the linear map
$S({X,Y}):=F(f)$ is  called the {\it amplitude} assigned to $f$.

\

Next, we would like to introduce a topology on the space of quantum
Riemannian $d$-spaces specified by the definition of the convergence.
The latter depends on the notion of convergence of Hilbert vector spaces which carry
self-adjoint positive operators (each operator is a generator of the semigroup
corresponding to the intervals $[0,t]$).
Let us discuss this notion of convergence.
Suppose that we have a  sequence of Hilbert spaces $H_{\alpha}$ parametrized
by $\alpha\in \Z_+\cup \{\infty\}$, where $H_{\infty}:=H$. Assume that
every $H_{\alpha}$ carries a self-adjoint non-negative (in general unbounded) operator $L_{\alpha}$ where $L_{\infty}:=L$.
We would like to define
a topology on the union of $H_{\alpha}$. We will do that following [KS].

\begin{defn} \label{convergence} 1) We say that a sequence $H_{\alpha}$ converges to
$H$ as $\alpha\to \infty$, if there exist an open dense subspace $D\subset H$, and
for any $\alpha$ there is a continuous linear map $R_{\alpha}:D\to H_{\alpha}$,
such that $lim_{\alpha\to \infty}||R_{\alpha}(x)||_{H_{\alpha}}=||x||_H$ for any
$x\in D$.

2) Suppose that $H_{\alpha}$ converges to $H$ in the above sense.
We say that a sequence $x_{\alpha}\in H_{\alpha}$ strongly converges to $x\in H$ if there is
a sequence $y_{\beta}\in D$ converging to $x$ such that

$$lim_{\beta\to \infty}lim_{\alpha\to \infty}||R_{\alpha}(y_{\beta})-x_{\alpha}||_{H_{\alpha}}.$$

3) We say that a sequence $(H_{\alpha}, L_{\alpha})$ strongly converges to $(H,L)$ if
$D$ contains the domain of $L$, the sequence $H_{\alpha}$ converges to $H$ is the sense of 1),
and for any sequence $x_{\alpha}\in H_{\alpha}$ which strongly converges to $x\in H$
and any continuous function $\varphi: [0,\infty)\to \R$ with compact support
the sequence $\varphi(L_{\alpha})(x_{\alpha})$ strongly converges to $\varphi(L)(x)$.

4) We say that a sequence of quantum Riemannian $d$-spaces
$\{M_{\alpha}\}_{\alpha\in \Z_+}$
converges to a quantum Riemannian $d$-space $M$ as $\alpha\to \infty$, if for any morphism
$f:X\to Y$ in $Riem_d^{NC}$ the corresponding sequence
$(H_{X,\alpha}^{\ast}\otimes H_{Y,\alpha}, S({X,Y,\alpha}))$ strongly converges to
$(H_X^{\ast}\otimes H_Y, S({X,Y}))$.

\end{defn}

\begin{rmk} \label{locally convex case}One can define the notion of quantum locally convex $d$-geometry
over a complete normed field $K$ by considering monoidal functors to the monoidal category
$Vect_K^{lc}$ of locally convex topological spaces over $K$. Probably the case of locally
compact $K$ (e.g. the field of $p$-adic numbers) is of some interest.

\end{rmk}

A natural generalization of the above definitions gives rise to the notion of {\it quantum
Riemannian $d$-space with a measure}. It is an analog of the notion of Riemannian manifold $M$ equipped
with a probability measure, which is absolutely continuous with respect to the measure $dvol_M/vol(M)$ defined by the Riemannian metric.
Namely, we assume that an additional datum is given: a continuous linear functional $\tau: H\to \C$
such that $\tau(e^{-tL}x)=\tau(x)$ for any $x\in H, t\ge 0 $, and which satisfies the  ``trace" property described below.

Let $\Gamma_{l_1,l_2,l}^{in_1,in_2, out}$ be a $Y$-shape graph with two inputs (marked
by $in_1,in_2$), one output (marked by $out$) and three edges with the lengths $l_i$
(outcoming from $in_i, i=1,2$) and $l$ (incoming to $out$). Then the trace property says:
$$\tau\circ S({\Gamma_{l_1,l_2,l}^{in_1,in_2, out}})=\tau\circ S({\Gamma_{l_2,l_1,l}^{in_2,in_1, out}}).$$
This is an analog of the property: $\tau(fg)=\tau(gf)$ in case when $H$ is obtained
by the Gelfand-Naimark-Segal construction from
a $C^*$-algebra  and a tracial state $\tau$ on it.

We define the category $Riem_d^{NC,mes}$ of {\it measured $(d-1)$-dimensional
Riemannian germs} with the objects which are $(d-1)$-Riemannian
germs as before, equipped with  probability measures which are absolutely continuous
with respect to the volume measure associated with the germ of the Riemannian metric. Morphisms between measured $(d-1)$-Riemannian germs
are measured compact Riemannian $d$-dimensional manifolds, such that in the neighborhood of a boundary
both metric and measure are trivialized as products of the given metric and measure
on the boundary and the standard metric and measure on the interval.

Then we modify the above Property 4)  by the following requirement:

if $x_{\alpha}\to x, x_{\alpha}\in H_{\alpha}, x\in H$, then $\tau_{\alpha}(x_{\alpha})\to \tau(x)$ for the corresponding
linear functionals. In this way we obtain a topology on quantum measured
Riemannian $d$-spaces.

In the above discussion of convergence we spoke about convergence of quantum Riemannian spaces, not underlying germs of Riemannian manifolds.
In other words, having a sequence of quantum Riemannian $d$-spaces given
by functors $F_{\alpha}: Riem_d^{NC}\to Hilb_{\C}, \alpha\in \Z_+$ we say that the sequence converges
to a quantum Riemannian $d$-space given by a functor $F:Riem_d^{NC}\to Hilb_{\C}$
if for any morphisms $f:X\to Y$ in $Riem_d^{NC}$ we have $F_{\alpha}(f)\to F(f)$ as
$\alpha\to \infty$. The latter convergence is defined in the Definition 4.2.1.
On the other hand, we have Gromov-Hausdorff (or measured Gromov-Hausdorff)
topology on the objects of $Riem_d^{NC}$ (we leave as an exercise to the reader to work out
the modification of either topology which takes into account labelings of the boundary
components). Moreover we have those topologies on the spaces of morphisms as well.
This can be used to define the notion of {\it continuous} functor.

\begin{defn} \label{continuous functor}We say that a functor $F:Riem_d^{NC}\to Hilb_{\C}$ is continuous
if for any sequence of morphisms $f_{\alpha}: X_{\alpha}\to Y_{\alpha}$ in
$Riem_d^{NC}$ which converges in the Gromov-Hausdorff sense to $f:X\to Y$
we have: $F(f_{\alpha})$ converges to $F(f)$ in the sense of Definition 4.2.1.

\end{defn}

In particular, if $F$ is continuous then $X_{\alpha}\to X, \alpha\to \infty$ implies
$F(X_{\alpha})\to F(X), \alpha\to \infty$.

\begin{rmk}\label{collapsing and non-collapsing continuity}
Continuity allows some dramatic changes in the limiting geometry, e.g. contraction of loops in the $d=1$ case. In order to avoid this problem one can use another version of continuity which
is  called {\it non-collapsing continuity}. It is defined in the same as  in the above
definition with the restriction that for the limit $f_{\alpha}\to f$ we require $vol(f_{\alpha})(B(0,r))\ge const>0$ for any ball $B(0,r)\subset f_{\alpha}$
of the Riemannian $d$-bordism $f_{\alpha}$. In particular, for $d=1$ non-collapsing continuity prohibits
contraction not only loops of a graph but edges as well.
\end{rmk}
There is a natural modification of the above definitions
to the case of quantum measured Riemannian $d$-spaces, which we leave as an
exercise to the reader. In this case measured Gromov-Hausdorff convergence of objects
in $Riem_d^{NC,mes}$ implies the above-mentioned convergence of Hilbert spaces and linear functionals.
More realistic continuity property corresponds to the limit
$\varepsilon\to 0$ for an $\varepsilon$-neighborhood of a submanifolds.

\section{Graphs and quantum Riemannian
$1$-geometry}\label{graphs and quantum 1-geometry}

The content of this section is heavily influenced by conversations
with Maxim Kontsevich. The notion of commutative Riemannian $1$-space is basically what
he called Graph Field Theory.

\subsection{Quantum Riemannian $1$-spaces}\label{quantum Riemannian 1-spaces}

A quantum Riemannian $1$-space is defined by the following data:

1) A class $G$ of {\it metrizable labeled graphs} $(\Gamma,I_{-},I_{+},l,p)$  described in 1a), 1b) below:

1a) $\Gamma\in G$ is a finite  graph with
external vertices having the valency one and labeled by elements
of the disjoint union of finite sets: $I=I_{-}\sqcup I_{+}$ (the letter $p$ above
denotes the labeling). Vertices
parametrized by the elements of $I_{-}$ (resp. $I_{+}$) are called {\it incoming} (resp. {\it outcoming}), or, simply, $in$ (resp. $out$)
vertices. We denote by $V_{in}(\Gamma)$ the set of inner vertices of $\Gamma$,
by $E(\Gamma)$ the set of edges of $\Gamma$, etc.

1b) A length function $l:E(\Gamma)\to \R_{>0}$ , where  $E(\Gamma)$
is the set of edges of $\Gamma$.

2) Separable real Hilbert space $H$ (or a complex Hilbert space with real structure).\footnote{As in in the case of CFT we can relax
this condition assuming that $H$ is a locally compact, in particular, a nuclear space. In case if $H$ is a complex vector space the formulas below have to be modified in order to include complex conjugation.}

3) To each $(\Gamma,l,I_{-},I_{+})$ a trace-class operator
$S(\Gamma,l,I_{-},I_{+}): \otimes_{I_{-}}H\to \otimes_{I_{+}}H$, which
we will often denote simply by $S(\Gamma)$.

These data are required to satisfy the following axioms:

QFT 1) If $\Gamma$ is obtained by gluing $\Gamma_1$ and $\Gamma_2$
(with an obvious definition of the sets $I_{\pm}$ and the length function)
then $S(\Gamma)=S(\Gamma_1)\circ S(\Gamma_2)$ (composition
of operators).

QFT 2) Operators $S(\Gamma)$ are invariant with respect to  isometries.

QFT 3) If $\Gamma^{\vee}$ is obtained from $\Gamma$ by relabeling so that
all incoming vertices are declared outcoming and vice versa then
$S(\Gamma^{\vee})=S(\Gamma)^{\ast}$ (conjugate operator). Here we use the scalar
product in order to identify $H$ and $H^{\ast}$.

We will add two more axioms below. They will allow us to compactify various
``moduli spaces" of quantum Riemannian $1$-spaces.

QFT 4) The operators $S(\Gamma)$ enjoy the non-collapsing continuity property.
In other words, a small  deformation of a metrized graph $\Gamma$ leads to a small change
of $S(\Gamma)$ in the {\it normed} operator topology.
In addition, the operator $S(\Gamma)$ is a continuous function (in the {\it strong} operator topology) with respect
to the length of an edge which is not a loop. Moreover if $F\subset \Gamma$ is a subforest
(i.e. a collection of internal edges without loops), and $(\Gamma/F,I_{-},I_{+},l)$ is the metrized graph obtained by contracting all of the edges from $F$, then
$S(\Gamma/F)$ coincides with the limiting operator  $lim_{l(F)\to 0}S(\Gamma)$, which is the
limit of $S(\Gamma)$ as lengths of all
edges belonging to $F$ simultaneously approach $0$.

 QFT 5) If $\Gamma_{\varepsilon}^0$ is a graph, which is a segment $[0,\varepsilon]\subset \R$
such that $I_{-}=\{0\}$ and $I_{+}=\{\varepsilon\}$ then
$lim_{\varepsilon\to 0}S(\Gamma_{\varepsilon}^0)=id_H$, where
the limit is taken in the strong operator topology.
Similarly, if $\Gamma_{\varepsilon}^1$ is a graph, which is
the same segment but with $I_{-}=\{0,\varepsilon\}$ and $I_{+}=\{\emptyset\}$, then
$lim_{\varepsilon\to 0}S(\Gamma_{\varepsilon}^1):H\otimes H\to \R$
is the scalar product on $H$.

\begin{exa}\label{kernels} Let $M$ be a compact Riemannian manifold with the metric
$g_M$. We can normalize the Riemannian measure $d\mu=\sqrt{det\,g_M}\,dx$
so that $\int_Md\mu(M)=1$. Let $H=L^2(M,d\mu)$. Then with each $t>0$ we
associate a trace-class operator $P_t=exp(-t\Delta)$, where $\Delta$
is the Laplace operator on $M$. One has an integral representation
$(P_tf)(x)=\int_MG_t(x,y)f(y)d\mu(y)$, where $G_t(x,y)$ is the heat kernel.
To a graph $(\Gamma,l) \in G$ we assign the following function on $M^{I_{-}\cup I_{+}}$:
$$K_{(\Gamma,l)}((x_i)_{i\in I_{-}},(y_j)_{j\in I_{+}})=\int_{M^{V_{in}(\Gamma)}}d\mu^{V_{in}(\Gamma)}\prod_{e\in E(\Gamma)}G_{l(e)}(x,y),$$
where $d\mu^{V_{in}(\Gamma)}=\prod_{v\in V_{in}(\Gamma)}d\mu$ is the product measure on $M^{V_{in}(\Gamma)}$,
associated with $d\mu$. In other words, we assign a measure $d\mu$ to every internal vertex, the kernel
$G_{l(e)}(x,y)$ (propagator) to every edge and then integrate over all internal vertices. We obtain
a function whose variables are parametrized by  input and output vertices.
Finally, we define $S(\Gamma, l):\otimes_{I_{-}}H\to \otimes_{I_{+}}H $  as the integral operator with the kernel $K(\Gamma,l)$.

\end{exa}

We see that every  compact Riemannian manifold defines a quantum Riemannian $1$-space.
In fact it is a {\it commutative} Riemannian $1$-space, since the operators $S(\Gamma,l)$ are invariant
with respect to the action of the product of symmetric groups $S_{I_{-}}\times S_{I_{+}}$ on the set
$I_{-}\times I_{+}$
(and hence the product on the algebra of smooth functions $A\subset H$ defined by the $Y$-shape graph is commutative).

Let us restate the above example in terms more suitable for non-commutative
generalization. In order to define a collection of operators $S(\Gamma, l):\otimes_{I_{-}}H\to \otimes_{I_{+}}H $ it suffices to have tensors $K(\Gamma,l)\in (\otimes_{I_{-}}H^{\ast})\otimes (\otimes_{I_{+}}H)$ satisfying the composition property: $K(\Gamma_1\circ \Gamma_2, l_1\circ l_2)=
ev^{\otimes_{I_{+}}H}(K(\Gamma_1,l_1)\otimes K(\Gamma_1,l_2))$, where $ev:H^{\ast}\otimes H\to \C$
is the natural pairing, $\Gamma_1\circ \Gamma_2$ is the gluing operation (see QFT 1)), and $l_1\circ l_2$
is the length function obtained by the natural extension of $l_1$ and $l_2$ to $\Gamma_1\circ \Gamma_2$.

Suppose that to each vertex $v\in V(\Gamma)$ we assigned
a tensor $T_v\in \otimes_{i\in Star(v)}H$, where  $Star(v)$ is the set of adjacent
vertices (i.e. vertices $w\in V(\Gamma)$ such that $(w,v)$ is an edge), and to every edge $e\in E(\Gamma)$ with the endpoints $v_1, v_2$ we assigned a linear functional
$\tau_{e} :H\otimes H\to \C$. We define $T(\Gamma):=\otimes_{v\in V(\Gamma)}T_v$.
Let $\tau_{I_+}$ be the tensor product $\otimes_{e}\tau_e$ taken over all
edges which are not of the form $e=(v,w)$, where $w\in I_+$ (i.e. $e$ does not have
an endpoint which is an $out$ vertex). Then the element
$\tau_{I_+}(T(\Gamma))$ belongs to $\otimes_{i\in I_+}H$.

In the above example we thought of $G_{l(e)}(x,y)$ as of an element of the space
$H\otimes H$, where tensor factors are assigned to  the endpoints of $e$.
To such an edge we assigned the pairing
$\tau_e(f,g)=\int d\mu_xd\mu_yG_{l(e)}(x,y)f(x)g(y)$,
where $d\mu_x=d\mu_y=d\mu$.
We remark, that in order to define $T(\Gamma)$ it suffices to
have vectors $z_i\in H, i\in I_-\cup I_+$ and tensors $T_e\in H\otimes H$
for every edge $e=(v_1,v_2)\in E(\Gamma)$.

\subsection{Spectral triples and quantum Riemannian $1$-geometry}\label{relation to spectral triples}

As we explained in the Introduction, a spectral triple in the sense of Connes gives rise to a
quantum Riemannian $1$-geometry. Let us make this point more precise.
We will use a slightly different definition than in [Co1], since
we would like to axiomatize the Laplace operator
rather than the Dirac operator as in the loc.cit. 

More precisely, let us consider
a triple $(A,H,L)$ which consists of an unital complex $\ast$-algebra $A$,
a separable Hilbert left $A$-module $H$ (i.e. the $\ast$-algebra $A$ acts on $H$ by bounded operators),
a self-adjoint non-negative unbounded operator $L$ on $H$, such that
the operator $P_t=exp(-tL): H\to H$ has finite trace for all $t>0$.
The latter implies that that spectrum of $L$ consists of eigenvalues only,
and the operator $(1+L)^{-1}$ is compact.

In order to  formulate the second assumption, for any $a\in A$ let us consider the function
$\phi_a(t)=e^{ta}Le^{-ta}, t\ge 0$. It takes value in unbounded operators in $H$,
and the domains of all $\phi_a(t)$ belong to the domain of $L$ for all $a\in A, t\ge 0$.
We say that $\phi_a(t)$ has $k$-th derivative at $t=0$ if it can be represented
as a sum $\phi_a(t)=C_0(a)+C_1(a)t+{C_2(a)\over {2!}}t^2+...+{C_k(t,a)\over {k!}}t^k$ such that $C_i(a)$ are
some (possibly unbounded, but with a non-empty common domain) operators in $H$ and $C_k(0,a):=C_k(a)$ is a bounded operator in $H$. We will denote it by $ad_a^k(L)$. Now the second assumption says that
$ad_a^2(L)$ exists for all $a\in A$. Basically this means that the double
commutator $[[L,a],a]$ exists and bounded for all $a\in A$.

Let us assume that such a spectral triple has finite metric dimension $n$ (a.k.a. $n$-summable,
see [Co1], [CoMar]) and satisfies the regularity conditions (see loc cit.) such that
the volume functional
$$\tau(f)=Tr_{\omega}(fL^{-n/2})$$
is finite for all $f\in exp(-tL)(H)$ (here $Tr_{\omega}$ denotes the Dixmier trace, see loc. cit.).
For simplicity we will also assume that $H$ is obtained from $A$ by a GNS completion
with respect to the scalar product $\tau(fg^{\ast})$ (this assumption can be relaxed).
{\it Then it looks plausible that one can associate to it a quantum Riemannian $1$-geometry
similarly to  what we did in the end of Section \ref{quantum Riemannian 1-spaces}.}

\section{Ricci curvature, diameter and dimension: probabilistic and spectral approaches to precompactness}\label{probablity and precompactness}

In this Section we review some results presented in [BBG], [Ba], [Led], [LV],
[St], [KS], [KMS], [KaKu1-2] (see also [U]). 

Recall that there are basically three
approaches to precompactness of metric (and metric-measure) spaces with the diameter
bounded from above and the Ricci curvature bounded from below.

1) ``Geometric" approach
(which goes back to Gromov, see e.g. [Gro1], and which was developed in a deep and non-trivial
way by Cheeger, Colding, Fukaya and many others, see e.g. [Fu], [ChC1-3])
deals with Gromov-Hausdorff (or measured Gromov-Hausdorff) topology, and (very roughly speaking)
embeds a compact metric space $(X,d)$ into the Banach space $C(X)$ of continuous functions
via $x\mapsto d(x,\bullet)$. Then precompactness follows from a version of Arzela-Ascoli theorem,
since the space $X$ is approximated by a finite metric space. By the nature of this approach one needs
the notion of Ricci curvature to be defined ``locally", in terms of points of $X$.

2) In the ``spectral" approach (see e.g. [BBG], [KaKu1-2]) one embeds the metric-measure space $(X,d,d\mu)$ into
$L_2(X,d\mu)$ via $x\mapsto K_t(x,\bullet)$ where $K_t(x,y)$ is the ``heat kernel" (which needs to be defined
if $(X,d)$ is not a Riemannian manifold).
Then precompactness follows from a version of Rellich's theorem, since the assumptions on the diameter
and curvature  imply that the image of $X$ belongs to a Sobolev space, which is compactly embedded
in $L_2(X,d\mu)$.
Precompactness relies on the estimates for eigenvectors and eigenvalues of the ``generalized
Laplacian". The former are still local while the latter depends on the global geometry of $X$.
The spectral topology in general does not coincide with the measured Gromov-Hausdorff topology
(restrictions on the diameter and Ricci curvature can make these topologies equivalent,
see e.g. [KS], Remark 5.1).

3) In the  ``probabilistic" approach (see a good review in [L], or original proofs in [LV], [St]) one uses the ideas of optimal transport (different point of view is presented in [AGS]). It is
a mixture of the previous two approaches, since one  studies
the ``heat flow" on the space $P(X)$ of probabilistic Borel measures on $X$, but
proves the precompactness theorem in the measured Gromov-Hausdorff topology via a kind of Arzela-Ascoli
arguments. The point is that the heat flow on measures can be interpreted either as a gradient
line of a functional (entropy) or as a geodesic for some metric on $P(X)$.
More precisely the  space $P(X)$ carries a family $W_p, p\ge 1$ of the so-called Wasserstein metrics.
In the case $p=1$ such a metric (called also Monge-Kantorovich metric ) being restricted to delta-functions $\delta_x, x\in X$ reproduces
the original distance on $X$. The distance $W_1(\phi,\psi)$ coinsides with the ``non-commutative distance"
on the space of states on $C(X)$ discussed in the Introduction, thus making a connection with
Connes's approach to non-commutative Riemannian geometry.
In order to use the above ideas of optimal transport on needs
a non-commutative analog of the Wasserstein metric $W_2$. This is an interesting open problem.
\footnote{ I thank to Dima Shlyakhtenko who pointed me out the paper [BiVo] where the non-commutative
analogs of the Wasserstein metrics $W_p$ were introduced in the framework of the free probability 
theory. Since we define the tensor product of quantum spaces by means of the tensor product
of algebras, rather than their free product, it is not clear how to use that definition for the
purposes of quantum Riemannian $1$-geometry.}

\subsection{Semigroups and curvature-dimension inequalities}\label{semigroups and CD inequalities}

We follow closely [Ba], [Led].

Let $(X,d\mu)$ be a space with a measure $d\mu$ (which is assumed to be a Borel
probability measure), and $P_t,t\ge 0$ a semigroup of bounded operators
acting continuously in the operator norm topology on the Hilbert space of real-valued functions $L_2(X,d\mu)$, and such that
$$(P_tf)(x)=\int_XG_t(x,y)f(y)d\mu,$$
where the kernels $G_t(x,y)d\mu$ are non-negative for all $t\ge 0$.
It is also assumed that $P_t(1)=1$, which is true for semigroups arising
from Markov processes (main application of loc.cit).

\begin{exa} \label{Brownian motion}For the Brownian motion in $\R^n$ starting from the origin one has
$$G_t(x,y)d\mu={1\over{(2\pi t)^{n/2}}}e^{-|x-y|^2/2t}dy.$$

\end{exa}

One defines the generator of the semigroup $P_t$ as
$$Lf=lim_{t\to 0}{(P_tf-f)\over t}.$$
Then on the domain of the  non-negative operator $L$ one has
${\partial\over{\partial t}}P_t(f)=LP_t(f).$ For the Brownian motion the operator $L$
is just the standard Laplace operator on $\R^n$.
We will assume that $L$ is symmetric on its domain (this corresponds to the so-called time reversible measures). This also implies that the finite measure
$d\mu$  is invariant with respect to the semigroup $P_t,t\ge 0$.
We will also assume that the domain of $L$ contains a dense  unital subalgebra
$A$ (typically, the algebra of real-valued smooth functions on a manifold,
or the algebra of real-valued Lipschitz functions on a metric space).
Following Bakry we introduce a sequence of bilinear forms
$A\otimes A\to A$:

0) $B_0(f,g)=fg,$

1) $2B_1(f,g)=LB_0(f,g)-fL(g)-L(f)g=L(fg)-fL(g)-L(f)g,$

2) $2B_n(f,g)=LB_{n-1}(f,g)-B_{n-1}(f,Lg)-B_{n-1}(Lf,g), n\ge 2$.

In the case when $L=\Delta$ is the Laplace operator in $\R^n$
(Brownian motion case) one has
$B_2(f,f)(x)=|Hess(f)|^2(x):=\sum_{1\le i,j\le n}(\partial^2 f/\partial x_i\partial x_j)^2(x).$
If $L$ is the Laplace operator on the Riemannian manifold $M$ then
$$B_1(f,f)=g_M(\nabla f, \nabla f),$$
$$B_2(f,f)=Ric(\nabla f,\nabla f)+|Hess(f)|^2,$$
where $Ric$ denotes the Ricci tensor. In the case of general Riemannian manifold
the Hessian matrix $Hess(f):=\nabla^2(f)$ can be defined as a second derivative
of $f$ in the Riemannian structure,
so the above equality for $B_2(f,f)$ still holds. All that can be axiomatized such as follows.

Suppose that we are given a triple $(A,H,L)$ such that

1) $H$ is a separable real Hilbert space;

2) $L$ is a (possibly unbounded) operator on $H$, which is symmetric
on its domain;

3) $A\subset H$ is a unital real algebra, dense in the domain of $L$, such that
$L(1)=0$.

\begin{defn}\label{CD condition} a) An operator $L$ satisfies a curvature-dimension condition
$CD(R,N)$, where $R\in \R,N\ge 1$ if for all $f\in A$ one has
$$B_2(f,f)\ge RB_1(f,f)+{(Lf)^2\over N}.$$

b) An operator $L$ has the Ricci curvature greater or equal than $R$ if it satisfies
$CD(R,\infty)$, i.e.
$$ B_2(f,f)\ge R\,B_1(f,f),$$
for any $f\in A$.
\end{defn}

\begin{rmk} \label{generalization to complex Hilbert spaces}The above considerations can be generalized to the case of complex Hilbert spaces
with the real structure defined by an involution $x\mapsto x^{\ast}, x\in H$,
which is compatible with the involution on $A\subset H$.
In that case we define a sequence of bilinear forms $B_n(f,g^{\ast})$. In what follows we will
discuss for simplicity real Hilbert spaces.

\end{rmk}

Suppose that a  triple $(A,H,L)$ satisfies the above conditions 1)-3).

\begin{defn}\label{Ricii curvature bounded from below}
We say that $(A,H,L)$  has Ricci curvature greater
or equal than $R$ if the operator $L$ satisfies $CD(R,\infty)$.
\end{defn}

\begin{exa}\label{CD for Riemannian manifolds} a) If $d\mu=e^hdvol_M$, where $g=g_M$ is a Riemannian metric
on the $n$-dimensional manifold $M$, $dvol_M$ is the corresponding volume form, and $h$ is a smooth 
real function then
$L=\Delta+\nabla(h)\nabla$, where $\Delta$ is the Laplace operator associated with
the metric $g$. It was shown by Bakry  and \'Emery (see [BaEm]) that $CD(R,N), N\ge n$ for $L$ is equivalent to the following inequality of symmetric tensors:

$$ Ric\ge Hess(h)+Rg_M+{dh\otimes dh\over{(N-n)}}.$$
In particular, $CD(R,n)$ implies $h=const$, and hence $Ric\ge Rg_M$ everywhere.

b) If $L=(d/dx)^2-q(x)d/dx$ then $CD(R,N)$ is equivalent to

$$ q^{\prime}\ge R+{q^2\over{N-1}}.$$

\end{exa}

\begin{defn}\label{measured spectral triples}
A spectral triple $(A,L,H)$
is called {\it measured spectral triple} if we are given also Gelfand-Naimark-Segal state
$\gamma$ on $A$ which is invariant with respect to the semigroup $exp(-tL), t>0$.
\end{defn}

\subsection{Wasserstein metric and $N$-curvature tensor}\label{Wasserstein metric}

Here we recall definitions and results of [LV], [St].

Let  $f:\R_{\ge 0}\to \R$ be a continuous convex function, such that
$f(0)=0$. If $x^Nf(x^{-N})$ is convex on $(0,\infty)$, we will say
that the function $f$ is $N$-convex, where $1\le N <\infty$.
If $e^xf(e^{-x})$ is convex on $(-\infty,\infty)$ we will say
that the function $f$ is $\infty$-convex.
\begin{exa} \label{example of convex function}
a) The function $f_N(x)=Nx(1-x^{-1/N})$ is $N$-convex for $1<N<\infty$.

b) The function $f_{\infty}(x)=xlog\,x$ is $\infty$-convex.

\end{exa}

Let  $(X,d\mu)$ be a compact Hausdorff space equipped with a finite probability
measure, and $f$ be an arbirtary continuous convex function as above.
For any probability measure $d\nu=\rho\,d\mu$, which is absolutely continuous with respect to $d\mu$ we define the $f$-relative entropy of $d\nu$ with respect to $d\mu$ by the formula
$$E_{d\mu}^f(d\nu)=\int_Xf(\rho(x))d\mu$$
(one can slightly modify this formula
in order to include measures with a non-trivial singular part in the
Lebesgue decomposition).

Let now $(X,d)$ be a compact metric space. We define the $L_2$-Wasserstein metric (or, simply, the Wasserstein
metric, since we will not consider other Wasserstein metrics) on the space $P(X)$ of all Borel probability measures on $X$
by the formula

$$W_2(d\nu_1,d\nu_2)^2=inf\{\int_{X\times X}d^2(x_1,x_2)d\xi\},$$
where infimum is taken over all probability measures $d\xi\in P(X\times X)$ such that $(\pi_i)_{\ast}d\xi=d\nu_i, i=1,2$,
where $\pi_i$ are the natural projections of $X\times X$ to the factors.
Then $(P(X),W_2)$ becomes a compact metric space (in fact a length space, if $X$
is a length space),
and the corresponding metric topology coincides with the weak $\ast$-topology on
measures.

\begin{defn}\label{Ricci N-curvature}([LV], [St])

a) We say that the compact metric-measure length space
$(X,d,d\mu)$ has a non-negative $N$-Ricci curvature, $1\le N< \infty$ if
for all $d\nu_0,d\nu_1\in P(X)$ which have supports belonging to $supp(d\mu)$
there is a Wasserstein geodesic $d\nu_t, 0\le t\le 1$ joining $d\nu_0$
and $d\nu_1$ such that for all $N$-convex functions $f$ and
all $0\le t\le 1$ one has
$$E_{d\mu}^f(d\nu_t)\le tE_{d\mu}^f(d\nu_1)+(1-t)E_{d\mu}^f(d\nu_0).$$
In other words, the $f$-relative entropy $E_{d\mu}^f$ is convex
along a geodesic joining $d\nu_0$ and $d\nu_1$.

b) Given $R\in \R$ we say that $(X,d,d\mu)$ has $\infty$-Ricci curvature bounded below by $R$ if
for all $d\mu_0,d\mu_1$ as in part a) there is a Wasserstein geodesic $d\nu_t$ joining $d\nu_0$
and $d\nu_1$ such that for all $\infty$-convex functions $f$ and all $t\in [0,1]$ one has:
$$E_{d\mu}^f(d\nu_t)\le tE_{d\mu}^f(d\nu_1)+(1-t)E_{d\mu}^f(d\nu_0)- {1\over{2}}\lambda(f)t(1-t)W_2(d\nu_0,d\nu_1)^2,$$
and $\lambda=\lambda_R$ is a certain map from $\infty$-convex functions to $\R\cup \{-\infty\}$ defined
in [LV], Section 5. For $f=xlog\,x$ one can take $\lambda(f)=R$.

\end{defn}

Let now $N\in [1,\infty]$, and $d\mu=e^h\,dvol_g$ be a probability measure on a smooth compact connected Riemannian manifold $(M,g), dim\,M=n$,
associated with an arbirtary smooth real function $h$. One defines the Ricci $N$-curvature
$Ric_N$ such as follows (see [LV]):

a) $Ric_N= Ric-Hess(h)$, if $N=\infty$,

b) $Ric_N=Ric-Hess(h)- {dh\otimes dh\over {N-n}}$, if $n<N<\infty$,

c) $Ric_N=Ric-Hess(h)-\infty\cdot(dh\otimes dh)$, if $N=n$. Here by convention
$0\cdot \infty=0$.

d) $Ric_N=-\infty$, if $N<n$.

The following theorem was proved in [LV], [St].

\begin{thm} \label{criterion for non-negative Ricci N-curvature}
1) For $N\in [1,\infty)$ the measured length space $(M,g,d\mu)$ has non-negative $N$-Ricci
curvature iff $Ric_N\ge 0$ as a symmetric tensor.

2) It has $\infty$-Ricci curvature bounded below by $R$ iff $Ric_{\infty}\ge Rg$.

\end{thm}

Notice that in the above assumptions the condition $Ric_N\ge Rg$ is equivalent to Bakry's $CD(R,N)$ condition
for the operator $\Delta+\nabla(h)\nabla$.

One defines the measured Gromov-Hausdorff topology on the set of metric-measure spaces in the usual way: a sequence $(X_i,d_i,d\mu_i)$ converges to $(X,d,d\mu)$ if there are
$\varepsilon_i$-approximations $f_i:(X_i,d_i)\to (X,d)$ such that
$\varepsilon_i\to 0$ as $ i\to \infty$, which satisfy the condition that the direct images $(f_i)_{\ast}d\mu_i$ converge (in the weak topology) to $d\mu$.

\begin{thm} \label{precompactness for Riici  N-curvature}([LV])

a) For all $1\le N<\infty$ the set of length metric-measure spaces with non-negative $N$-Ricci curvature is precompact and complete in the measured Gromov-Hausdorff topology.

b) For $N=\infty$ the same is true for the set of length metric-measure spaces
with the $\infty$-Ricci curvature greater or equal than fixed $R$.

\end{thm}

Similar result was proved in [St], where
the precompactness theorem was established with respect to the following distance function on metric-measure spaces:

$${\bf D}^2((X_1,d_1,d\mu_1),(X_2,d_2,d\mu_2))=inf\{\int_{X_1\times X_2}d^2(x_1,x_2)d\chi(x_1,x_2)\},$$
where infimum is taken over all $d\chi \in P(X_1\times X_2)$ which projects
onto $d\mu_1$ and $d\mu_2$ respectively under the natural projections, and all
metrics $d$ on $X_1\sqcup X_2$ which coincide with the given
metrics $d_i, i=1,2$ on $X_i,i=1,2$.

\subsection{Remark about the Laplacian}\label{remark about Laplacian}

As we have seen, the notion of Laplacian (maybe generalized one)
plays an important role in the description of the collapsing CFTs.
In the case of metric-measure spaces one can use the following
approach to the notion of Laplacian (see [Kok]).
Let $(X,d,d\mu)$ be a metric-measure space. For any point
$x\in X$ and the open ball $B(x,r)$ with the center at $x$ we define
the operator ``mean value" $f\mapsto \lan f\ran_{B(x,r)}$, where
$$\lan f\ran_{B(x,r)}={1\over{\mu(B(x,r))}}\int_{B(x,r)}f(y)d\mu(y).$$
Let us assume that the measure of every ball is positive.
Then one defines the Laplacian $\Delta_{d\mu}(f)$ by the formula
$$\Delta_{d\mu}(f)(x):=lim_{r\to 0}sup_{r>0}{2\over{r^2}}\lan(f-f(x))\ran_{B(x,r)} =$$
$$lim_{r\to 0}inf_{r>0}{2\over{r^2}}\lan(f-f(x))\ran_{B(x,r)},$$
provided the last two limits exist and coincide. It was observed in
[Kok] that this {\it measured} Laplacian coinsides with ${1\over {n+2}}\Delta_g$
on a Riemannian $n$-dimensional manifold $(M,g)$. Moreover, for a large
class of metric-measure spaces it is symmetric non-negative operator
on certain classes of functions built out from Lipschitz functions on $(X,d,d\mu)$.
In a different framework of spaces with
diffusion and codiffusion the notion of the heat operator and the Laplacian
was in introduced in [Gro2], Sect.3.3. The Laplacian introduced
in [Gro2] is in fact a vector field, which is the gradient of the energy function.

{\it Probably in all  cases when the Laplacian or the heat operator can be defined as a symmetric
non-negative operator, there is an interesting quantum Riemannian
$1$-geometry of  spaces with non-negative Ricci curvature.}

\subsection{Spectral metrics}\label{spectral metrics}
Here we follow [BBG].

Let $(M,g_M)$ be a compact closed Riemannian manifold, $d\mu=dvol_M$ denotes
the associated Riemannian measure. Let us choose an orthonormal
basis $\psi_j, j\ge 0$ of eigenvectors of the Laplacian
$\Delta=\Delta_{g_M}$, which is unbounded non-negative self-adjoint operator on the Hilbert space $H=L_2(M,d\mu)$.
Let $\lambda_j$ be the eigenvalue corresponding to $\psi_j$.
Then for
every $t>0$ one defines a map $\Phi_t: M\to l_2(\Z_+)$ by the formula
$$\Phi_t(x)=\sqrt{Vol(M)}(e^{-\lambda_jt/2}\psi_j(x))_{j\ge 0}.$$
In this way one obtains an embedding of $M$ into $l_2(\Z_+)$. 

For any two Riemannian
manifolds $(M_1,g_{M_1})$ and $(M_2,g_{M_2})$ as above, and any two choices
of the orthonormal bases one can compute the Hausdorff distance $d_H$
between compact sets $\Phi_t(M_1,g_{M_1})$ and $\Phi_t(M_2,g_{M_2})$ inside of the
metric space $l_2(\Z_+)$. This number depends on the choices of
orthonormal bases for the Laplace operators on $M_1$ and $M_2$, but one can remedy the problem
by taking $sup\,\,inf \,d_H$ over all
possible pairs of choices. This gives the distance
$d_t((M_1,g_{M_1}),(M_2,g_{M_2}))$  defined in [BBG]. 

It was proved in the loc.cit.
that the distance (for a fixed $t>0$) is equal to zero if and only if
the Riemannian manifolds are isometric.
More invariant way to spell out this embedding is via the heat kernel.
Namely, one considers the map $M\to L_2(M,d\mu)$ such that
$x\mapsto K_M(t/2,x, \bullet)$, where
$K_M(t,x,y)=\sum_{j\ge 0}e^{-\lambda_j t}\psi_j(x)\psi_j(y)$ is the heat kernel.
Subsequently one can identify $L_2(M,d\mu)$ with $l_2(\Z_+)$ since any two separable Hilbert spaces
are isometric. Thus we obtain the above embedding.

\

Let ${\cal M}(n,R,D)$ be the set of compact closed Riemannian manifolds
such that $dim\,M=n, Ric(M)\ge R, diam(M)\le D$. It was proved in [BBG]
that the $\Phi_t$-image of ${\cal M}(n,R,D)$ belongs  to a
bounded subset of the Sobolev
space $h^1(\Z_+)$ (the latter consists of sequences $(a_0,a_1,...)\in l_2(\Z_+)$
such that $\sum_{j\ge 0}(1+j^{2/n})a_j^2<\infty$). By Rellich's theorem the
embedding $h^1(\Z_+)\to l_2(\Z_+)$ is a compact operator. This implies
that the image of the embedding of ${\cal M}(n,R,D)$ via $\Phi_t$ is precompact in
$l_2(\Z_+)$. Moreover, the eigenvalues of the Laplacian are continuous
with respect to the spectral distance $d_t$.
Since only smooth manifolds were considered in [BBG] the measure $d\mu$ was
always the one associated with the Riemannian metric. Approach of [BBG]
was further developed and generalized in [KaKu1-2]. In the loc. cit the authors
discussed the compactification of ${\cal M}(n,R,D)$ with respect to their
version of the spectral distance, which is different from the one in [BBG].

\subsection{Spectral structures and measured Gromov-Hausdorff topology}\label{measured GH topology}

Here we briefly recall the approach suggested in [KS].

As we have seen above, the
Laplacian and the  heat  kernel can be defined for more general spaces than
just Riemannian manifolds (see e.g. [KMS], [S]
for the case of Alexandrov spaces). The notion of spectral structure was
introduced in [KS] with the purpose to study
the behavior of eigenvalues of the Laplacian with respect to perturbations of the metric
and topology of not necessarily compact Riemannian manifolds.

\

A {\it spectral structure} is a tuple of compatible data: $(L,Q,E,U_t,R(z),H)$ which consists of a separable
Hilbert space $H$ (complex or real), self-adjoint non-negative linear operator $L:H\to H$,
densely defined quadratic form $Q$ generated by $\sqrt{L}$, spectral measure $E=E_{\lambda}(L)$, pointwise continuous contraction semigroup $U_t$ with the infinitesimal generator $L$,
pointwise continuous resolvent $(z-L)^{-1}$. One can associate a spectral structure with
a pointed locally compact metric space equipped with a Radon measure (or with
a not necessarily pointed  compact metric-measure space). Two types of topologies
on the set of spectral triples were introduced in [KS]: strong topology and compact topology.
For both topologies the natural forgetful map from ``geometric" spectral structures to the
corresponding metric-measure spaces is continuous.

Main results of [KS] concern
convergence of spectral structures under the condition that the underlying metric-measure
spaces converge. In particular, one has such a convergence for the class of Riemannian
complete (possibly non-compact) pointed $n$-dimensional manifolds with the Ricci curvature
bounded from below ([KS], Theorem 1.3).
The results of [KS]
can be considered as  a generalization of the results of [Fu], [ChC1] about continuity
of eigenvalues of the Laplacian with respect to measured Gromov-Hausdorff topology.
Every commutative Riemannian $1$-space associated with a complete pointed Riemannian
manifold (or compact closed non-pointed Riemannian manifold) gives rise to a spectral structure. Therefore, applying results of [KS] one can deduce
precompactness of the moduli space of such Riemannian $1$-spaces.
Unfortunately, [KS] does not contain any precompactness results about the moduli space
of ``abstract" spectral structures, i.e. those which are not associated with meatric-measure spaces. Same is true for [BBG].

\subsection{Non-negative Ricci curvature for quantum $1$-geometry}\label{non-negative Ricci in 1-geometry}

Suppose that we are given a quantum Riemannian $1$-space which satisfies the following property:
there is a dense pre-Hilbert subspace $A\subset H$,
such that for any graph $\Gamma\in G$ the tensor product $A^{\otimes I_{-}}$ belongs
to the domain of the operator $S(\Gamma)$. Then, taking a $Y$-shape graph, as in the
Introduction, we recover from the axioms QFT 1)-QFT 4) an associative product on $A$.
Following Kontsevich\footnote{private communication} we impose the following three conditions:

 QFT 6) (spectral gap, or boundness of the diameter) Let $L$ be a generator of the semigroup $S(\Gamma^0_{\varepsilon})$ (see QFT 5)).
Then the spectrum of $L$ belongs to the set $\{0\}\cup \{a\}\cup [b,+\infty)$, where
$b>a>0$ are some numbers.

QFT 7) (7-term relation) For a graph $\Gamma(l_1,l_2,l_3,l_4)$ with one internal
vertex and four attached edges of length $l_i, 1\le i\le 4$ one has the following identity:

$$\sum_{1\le i\le 4}{\partial\over{\partial l_i}} S_{\Gamma(l_1,l_2,l_3,l_4)}=\sum_{1\le j\le 3}{\partial\over{\partial\varepsilon_j}}\arrowvert_{\varepsilon_j=0} S_{\Gamma(l_1,l_2,l_3,l_4,\varepsilon_j)},$$
where $\Gamma(l_1,l_2,l_3,l_4,\varepsilon_j)$ is a graph obtained from $\Gamma(l_1,l_2,l_3,l_4)$
by inserting one internal edge of the length $\varepsilon_j$ (three summands in the RHS correspond
to three different ways of pairing of four external vertices of the graph $\Gamma(l_1,l_2,l_3,l_4)$,
see the Figure).

\vspace{2mm}

%FIGURE 1 (corolla with 4 external vertices and 3 graphs obtained from it by inserting a new
%internal edge)
\begin{figure}
%\centerline{\epsfbox{Figure-1-7-term-relation.eps}}
%\centering
%\scalebox{0.5}
{\includegraphics{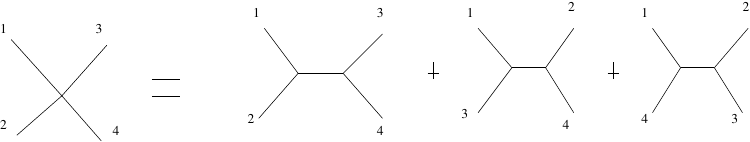}}
\end{figure}
\vspace{2mm}

QFT 8) (non-negativeness of Ricci curvature) If $\Gamma(l_1,l_1,l_3,l_3,\varepsilon)$ is the
graph $\Gamma(l_1,l_2,l_3,l_4,l)$ from QFT 7) with $l_2=l_1, l_4=l_3,\varepsilon=l$ then the following identity holds:

$$(\partial/\partial l_1-\partial/\partial l)^2S_{\Gamma(l_1,l_1,l_3,l_3,l)}\ge 0.$$

This condition is equivalent to the following one:

$$({d\over{dt}})^2|_{t=0}(e^{tL}(e^{-tL}(f)\cdot e^{-tL}(f)))\ge 0,$$
for all $f\in A$. The LHS of this inequality is equal to $B_2(f,f)$. Another way to state the above
inequality is to say that the amplitude $S_{\Gamma(l_1+t,l_1+t,l_3,l_3,l-t)}$ is convex at $t=0$.
The continuity of the amplitudes $S_{\Gamma, l}$ (see QFT 4)) implies that the condition $B_2(f,f)\ge 0$
is preserved if we contract a subforset of $\Gamma$ (i.e. we allow to contract a disjoint union of trees, while the contraction
of loops is prohibited).

\begin{rmk} \label{identity for Laplace}
a) The condition QFT 6) is motivated by the property that for any $t>0$ we have $Tr(e^{-tL})<\infty$.
This implies that there is an interval $(0,\lambda), \lambda>0$ which contains finitely many points of the spectrum of $L$.
If $L=\Delta$, the Laplace operator on a compact Riemannian manifold $M$, then the spectral gap $\lambda$
is of the magnitude $1/(diam M)^2$.

b) The condition QFT 7) is motivated by the case $L=\Delta$. In this case
$$\Delta(f_1f_2f_3)-\Delta(f_1f_2)f_3+...+\Delta(f_1)f_2f_3-\Delta(1)f_1f_2f_3=0$$
(7-term relation for the second order differential operator $\Delta$).

c) The condition QFT 8) is motivated by the $CD(0,\infty)$ inequalities of Bakry.

It was shown by Kontsevich that Segal's axioms of the unitary CFT imply that the
collapsing sequence of unitary CFTs gives rise to a commutative Riemannian $1$-space,
which satisfies axioms  QFT1)-QFT 8). The underlying Riemannian manifold $M$ is
the smooth part of  $X=Spec(H^{small})$ (see Section 2.2) with $L$ derived from
the limit of the rescaled Virasoro operators $(L_0+\overline{L}_0)/\varepsilon$, where
$\varepsilon$ is the minimal eigenvalue of $(L_0+\overline{L}_0)$, $\varepsilon\to 0$.

\end{rmk}

Recall that for a non-negative self-adjoint operator $L$ acting in a Hilbert space $H$
a {\it spectral gap} $\lambda_1(L)$  is the smallest positive eigenvalue of $L$.
Let us consider quantum Riemannian $1$-spaces with measure, for which the Hilbert space $H$
is obtained by the Gelfand-Naimark-Segal construction from a unital involutive
algebra $A$ and a state $\tau:A\to \C$. To every such a quantum space one can associate a spectral triple $(A,H,L)$.

\begin{conj} The space of isomorphism classes of quantum Riemannian $1$-spaces\label{conjecture about precompactness of quantum spaces}
with a measure, which have non-negative
Ricci curvature, spectral gap bounded below by a given number $C$, and such that the corresponding spectral
triples  have dimensional spectrum belonging to a given interval $[a,b]$ (see [CoMar] for the definition
of the dimension spectrum) is precompact in the topology defined in Section \ref{spaces with measure}.

\end{conj}

\

Finally, we are going to discuss an example in which a version of the above conjecture was verified.
Let $M_j, j\ge 1$ be a sequence of compact Riemannian manifolds of the same dimension $n$ with the diameter equal to $1$ (recall that rescaling of the metric does not change the Ricci curvature). Then, following Example \ref{kernels} we can associate with this sequence a sequence
of non-commutative Riemannian $1$-spaces $V(M_j), j\ge 1$. Suppose that $N$ is a measured Gromov-Hausdorff limit of $M_j$ as $j\to \infty$. Then it follows from [ChC3] that $N$ carries a generalized Laplacian,
and the measure, hence the same formulas as in Example 5.1.1 allows us to associate with $N$ a quantum Riemannian $1$-space $V(N)$. Let us say that $V(M_j)$ {\it weakly converges} to $V(N)$
if for any sequence of Lipschitz functions $f_j: M_j\to \R, f:N\to \R$ such that $|f_j\circ \psi_j-f|_{L_{\infty}(N)}$ as $j\to \infty$, and for any metrized graph $\Gamma$, we have:
$S_{\Gamma}(f_j)\to S_{\Gamma}(f)$ as $j\to \infty$. Here
$\psi_j: N\to M_j$ is any sequence of $\varepsilon_j$-approximations such that $\varepsilon_j\to 0$
as $j\to \infty$. Weak convergence gives rise to the topology on the space of equivalence classes of
quantum Riemannian $1$-spaces of {\it geometric origin} (i.e. those which correspond
to metric-measured spaces).
Then the following result holds (see [En] for the proof).
\begin{thm} \label{precompactness in weak topology}
The subspace of the space of the above quantum Riemannian $1$-spaces corresponding to  manifolds
with non-negative Ricci curvature is precompact in the weak topology.

\end{thm}

Having in mind results of [LV] and [St] we expect that it is in fact compact.

\section{Appendix: Deformation theory and QFTs on metric spaces}\label{QFTs on metric spaces}

\

{\bf by Maxim Kontsevich and Yan Soibelman}

\

Here we reproduce a portion of the unfinished joint paper  with Maxim Kontsevich
with the title ``Deformations of Quantum Field Theories" started in December 2000 and abandoned soon in favor of more urgent projects.
Among other things the draft contains a definition of  QFT on an arbitrary
metric space-time. Hopefully for QFTs on  Riemannian manifolds,
this approach can be translated into the language similar to the one used in the main body of this paper. 
Then this kind of ``generalized QFTs" will
give  examples of  quantum  metric-measure spaces.

From the point of view of present paper
it is also natural to ask about the behavior of these generalized QFTs with respect to the
Gromov-Hausdorff topology on the ``moduli space" of  space-times
which have non-negative Ricci curvature. This is an open question.

\subsection{Moduli space of translation-invariant QFTs: what to expect?}\label{moduli of translation-invariant QFTs}

Here we will briefly explain part of the structures we would like
to have in the deformation theory of a general translation-invariant QFT
on ${\R}^d$.
For such a theory ${\cal C}$ we have a space of local
fields $H$, which is filtered
by dimensions $\Delta\in \R_{\ge  0}$. It is convenient
to introduce the space
$\Omega^{\bullet}(H)=H\otimes\wedge^{\bullet}({\R}^{\ast d})$
of $H$-valued differential forms. We introduce the grading in this
space, so that the degree of $dx_i, 1\le i\le d$ is $-1$.
Since translations act on $H$ we have the action of the
corresponding vector fields $\partial_i=\partial_{x_i},1\le i\le d$ on $H$.
Therefore $\Omega^{\bullet}(H)$ carries a differential $m_1=d$ (Koszul
differential). Intuitively the tangent space to formal deformations
of the QFT with the space of fields $H$ is given by
$H/\sum_i\partial_iH$. This corresponds to a part
$$H\otimes\wedge^1({\R}^{\ast d})\to H $$
of the complex $(\Omega^{\bullet}(H),m_1).$
We will explain why the deformation theory of ${\cal C}$
is controlled by a certain $L_{\infty}$-algebra structure
on $\Omega^{\bullet}(H)_{\le 0}$ (here $\le 0$ stands for
the total degree).
 This structure is related
to the action of $H\otimes\wedge^{d-1}({\R}^{\ast d})$ on
$\Omega^{\bullet}(H)$, and we are going to explain this
action.

Since $H$ is filtered, we can consider the corresponding
graded space. It gives rise to a CFT $gr({\cal C})$.
The renormalization group flow contracts ${\cal C}$ to
$gr({\cal C})$.

\

Next we come to a question: what is the moduli space of QFTs
(on a given space-time)? Without answering this question, one has
difficulties in defining deformations of a given QFT.
In practice physicists speak about ``deforming the Lagrangian
of a theory". This is not satisfactory, because Lagrangians
are not fundamental objects. One can have a theory (a CFT, for example)
which is a priori defined without a Lagrangian.

We suggest a point of view which can be explained by analogy with Morse
theory: one can reconstruct a compact manifold from a general gradient
field on it. Let us explain this approach in a ``simplified picture
of the world". Namely, imagine that there is a
 smooth moduli space ${\cal M}$ of translation-invariant QFTs
on a given Euclidean space-time  ${\R}^d$. This moduli space carries
a vector field (called $\beta$-function by physicists). This vector field
has the following origin. Let us consider the natural  action of the group $\R_{>0}$ on
the metric $g$ on ${\R}^d$ given by $g\mapsto \lambda \,g$.
We {\it assume} that there is a canonical lift of this action
to ${\cal M}$. The corresponding group is called the
{\it renormalizarion group} (RG for short), and by definition it acts on
quantum field theories. We prefer to call it RG vector field
(rather than $\beta$-function) but keep the notation $\beta$.
Locally it is a gradient vector field.

Critical points of the RG vector field $\beta$ are CFTs. We assume that
for each Morse critical point $x$ we have:
${\cal M}={\cal M}_x^{in}\bigcup {\cal M}_x^{out}$ (union of points
which are attracted to $x$ and repelled from $x$ as $t\to +\infty$).

\begin{defn} \label{renormalizable QFTs}
Renormalizable (with respect to $x$) QFTs correspond
to points of ${\cal M}_x^{out}$. Unrenormalizable QFTs correspond to
points of ${\cal M}_x^{in}$.

\end{defn}

Let us assume that $\beta$ is a gradient
vector field: $\beta=grad(c)$ locally near $x_0\in {\cal M}, \beta(x_0)=0$.
Here $c(x)$ is a smooth function on ${\cal M}$
(Zamolodchikov c-function), which near the Morse
critical point $x_0$ can be written as $c(x)=\sum_i\lambda_ix_i^2$
(possibly infinite sum).

Let us assume  that $Hess(c):=d^2(c)_{|T_{x_0}{\cal M}}$ has finitely many
negative eigenvalues, $\lambda_i=\Delta_i-d$. It follows that
${\cal M}_{x_0}^{out}$ is a finite-dimensional manifold.
Since $\beta$ is a gradient vector field, all numbers $\Delta_i$
(called dimensions of local fields) are real.
The deformation theory we described at the very beginning of this
section is the deformation theory of the CFT associated with the
point $x_0\in {\cal M}$ in the direction of ${\cal M}_{x_0}^{out}$.

All deformed QFTs can be described by means of
the operator product expansion (often abbreviated as OPE). There is an $L_{\infty}$-algebra
controlling these deformations. It gives rise to the finite-dimensional
moduli space ${\cal M}_{x_0}^{out}$. More precisely, the
$L_{\infty}$-algebra gives rise to a formal pointed dg-manifold
(see [KoSo2], [KoSo3]) with $x_0$ be the marked point.
It is foliated (in the sense of dg-manifolds) by leaves
of the odd vector field. Equivalent theories belong to the
same leaf. ``Moduli space" of leaves is isomorphic to ${\cal M}_{x_0}^{out}$.

\begin{rmk} \label{non-Morse critical points}
The above description is still not sufficiently general.
Critical points of $\beta$ are not necessarily Morse.
Linearization of $\beta$ is no longer acting on the tangent space
to the critical point without kernel. As a result one has
``marginal" local fields such that $\lambda_i=\Delta_i-d=0$.
The OPE can contain fractional powers.

\end{rmk}

The deformation theory we are going to develop deals
 with OPE rather than with the quantum field theory itself.
This means that a QFT is determined by correlators between
fields. The latter depend not only on OPE but also on
the boundary conditions at infinity. In this section we will
ignore the behavior at infinity.

\subsection{Renormalization}\label{renormalization}
Mathematical framework for renormalization was discussed in a series of papers by Connes and Kreimer
(see e.g. [CoKr]). Here we present a slightly different point of view, motivated
by our approach to deformation theory of QFTs.

The OPE is determined by the behavior of the correlators
when some of points collide. This means that the Fulton-Macpherson
operad $FM({\R}^d)$ (see [KoSo2]) can be useful.

For example, let us consider the space $H\otimes \Omega_c^d({\R}^d)$
of $H$-valued compactly supported differential forms of the top degree.
For a choice $\phi_1,...,\phi_n\in H\otimes \Omega_c^d({\R}^d)$
we can define the universal partition function $Z$ such that

$$Z(\varepsilon\sum_{1\le i\le n}\phi_i)=\sum_{n\ge 0}{\varepsilon^n\over n!}
\sum_{i_1,...,i_n}\int_{({\R}^d)^n}^{\prime}\lan\phi_{i_1}...\phi_{i_n}\ran,$$
where $\int^{\prime}$ denotes a regularization of the integral
at $x_i=x_j$, and $\lan....\ran$ denotes the correlator.
Notice that since we take compactly supported forms on ${\R}^d$
the regularized integral converges.

From this formula we see that
$$\lan\phi_{i_1}...\phi_{i_n}\ran={\delta^nZ\over \delta\phi_{i_1}...
\delta\phi_{i_n}}.$$

Morally the formulas above should correspond to the deformation
of a QFT by adding $\sum_i\phi_i$ to the Lagrangian.
The choice of the regularization is not fixed.
Normally people take as the correlators for the deformed
theory the coefficients for $\varepsilon^0$.
We can take any linear functional $R:C^{\infty}(0,r)\to {\R}$
such that if $f$ is continuous at zero then $R(f)=f(0)$.
We call such a functional a {\it regularization operator}.

Assume that we have fixed a regularization operator.
Then for a choice of fields
$\phi\in\varepsilon \Omega^{\bullet}(H)[[\varepsilon]],
\phi_i \in \Omega^{\bullet}(H),1\le i\le n$,
choice of points $x_i\in {\R}^d, 1\le i\le n$
we can introduce new correlators
$\lan\phi_1(x_1)...\phi_n(x_n)\ran_{\phi,R}
\in C^{\infty}(({\R}^d)^n\setminus diag)[[\varepsilon]]$
such that

$$\lan\phi_1(x_1)...\phi_n(x_n)\ran_{\phi,R}=\sum_{k\ge 0}{1\over k!}
R(\int d^ky\lan\prod_{1\le j\le k}\phi(y_j)
\prod_{1\le i\le n}\phi_i(x_i)\ran),$$
where the integration is taken over the subspace $|y_i-y_j|>\delta,
|y_i-x_j|>\delta, |y_i|<{1\over \delta}$, and $\delta>0$ is some
number.

\begin{defn} \label{deformed theories}
1) We say that the correlators $\lan...\ran_{\phi,R}$ define
the deformed theory.

2) Two deformed theories given by $(\phi,R)$ and
$(\phi^{\prime},R^{\prime})$
are called equivalent if there exists
$L\in End(H)[[\varepsilon]]$ that $L=id+o(1)$ and
$\lan\prod_i\phi_i(x_i)\ran_{\phi,R}=
\lan\prod_i(L\phi_i)(x_i)\ran_{\phi^{\prime},R^{\prime}}$
for any choice of fields $\phi_i$ and pairwise different points
$x_i$.
\end{defn}

\begin{rmk} \label{remarks about deformations of QFTs}
a) The field $\phi$ gives a tangent vector to the
space of deformations of the given QFT. It can be taken from
$H\otimes\wedge^d((\R^d)^{\ast})$.

b) The above definitions should be modified because ${\R}^d$ is non-compact.
We can either assume that the space-time is compact or introduce
a cut-off at infinity. For pedagogical reasons the reader can assume
that instead of $\wedge^{\bullet}(({\R}^d)^{\ast})$ we take the compactly
supported differential forms on ${\R}^d$, but in this case the 
translation invariance is lost.

c) The definition above corresponds to the following intuitive picture:
in order to deform a QFT we consider the space consisting of all Lagrangians
${\cal L}$ modulo $\partial_i{\cal L}, 1\le i\le d$ plus a regularization
procedure. A choice of an element in this space gives a QFT.
(Morally a choice of ${\cal L}$ gives an affine structure on the
space of QFTs). Then we need to identify all gauge equivalent theories.
The moduli space of QFTs consists of the classes of gauge equivalent
theories.

\end{rmk}

\subsection{The operad controlling  the OPE}\label{operad controlling OPE}

In order to explain the conjecture about $L_{\infty}$-structure we need
to discuss the role of Fulton-Macpherson operad $FM({\R}^d)$
in OPE. Traditionally, when talking about OPE, people
consider two colliding points in the space-time.
More generally one can consider several points approaching
to the same one with the same speed. It gives only one stratum
in the Fulton-Macpherson compactification of the space-time.
There are more strata, corresponding to groups of points colliding
to the same one
``with the same speed" (see [FM]). Hence the ``true" picture for OPE
should describe the asymptotic behavior of correlators
near all the strata. Each operadic space $FM_n({\R}^d)$ is a {\it real
manifold with corners}. We assume that  correlators
at each corner are functions having asymptotic expansions  given by  series
 $\sum_{\beta,n}r^{\beta}log^nrf_{\beta,n}(s), \beta \in {\Q},n\in {\Z}_+$, where $r$ is the
distance to the corner, and $f_{\beta,n}(s)$ is  a smooth function in all
other variables. Such functions behave nicely with respect to the
operadic composition in $FM({\R}^d)$. The space
of such functions on $FM_n({\R}^d)$ will be denoted by
$V_n$. We should also take care about two things:

a) filtration of $H$ by dimensions;

b) action of the group of translations.

Let ${\C}_{\varepsilon}$ be the space of formal series
$\sum_ic_i\varepsilon^{\beta_i}$ such that $c_i\in {\C}$
and $\beta_i\in {\R}, lim_{i\to +\infty}\beta_i=+\infty.$
We define the cooperad ${\cal A}=({\cal A}_n)_{n\ge 1}$
such that
${\cal A}_n^{\ast}={\C}_{\varepsilon}\widehat{\otimes}V_{n+1}
\widehat{\otimes}{\C}[[x]],$ where $\widehat{\otimes}$ means
the completed tensor product and the formal series ${\C}[[x]]$
is the space dual to the algebra of differential operators
with constant coefficients on ${\R}^d$ (it is responsible
for the action of the group of translations on $H$).
Notice that the spaces ${\cal A}_n^{\ast}$ are filtered by
the degrees $\beta_i$ and the powers of $r$ at the
corners.
The operadic structure on ${\cal A}$ comes
naturally from the operadic structure
on $FM({\R}^d)$.

\begin{conj} \label{conjecture about controlling algebra}
Let $\g$ be the $L_{\infty}$-algebra
which controls formal deformations
of $H$ as an ${\cal A}$-algebra. Then:

a) There is a structure of $L_{\infty}$-algebra on
$\Omega^d(H)=H\otimes\wedge^d({\R}^d).$

b) There exists an $L_{\infty}$-morphism $\Omega^d(H)\to \g.$

c) There is a structure of a $d$-algebra (see [KoSo2]) on
$\Omega^{\bullet}(H)$.

\end{conj}

In particular, it follows from c) that the cooperad of differential
forms on the configuration space of $d$-dimensional discs in $({\R}^d)^{\ast}$
acts on $\Omega^{\bullet}(H)$.

\subsection{Some metric geometry}\label{some metric geometry}

In this subsection we will introduce useful constructions and the language
which will be useful in the discussion of  QFTs on metric spaces.

Let $X=\{x_1,...,x_n\}$ be a finite set, $T$ be a planar tree with
the tails parametrized by $X$, and $\{\lambda_v\}, 0<\lambda_v<1$
be the set of numbers parametrized by all but tail vertices of $T$.

\begin{prp} \label{metric}
There exists a metric $\rho$ on $X$ such that
for any two different points $x_i, x_j$ one has

$$\rho(x_i,x_j)=C\prod_{v<i,j}\lambda_v,$$
where $C>0$ depends on the set $X$ only, and
the notation $v<i$ means that the vertex $v$ is
closer to the root vertex with respect to the natural
order on the vertices of $T$ (so that the root vertex is
the smallest element and tail vertices are maximal elements).

\end{prp}

\subsection{Clusters}\label{clusters}

Let $X$ be a finite set, $|X|\ge 2$,  equipped with a metric $\rho$.
We will assign a tree $T=T_X$ to these data.

\begin{defn} \label{definition of cluster}
Subset $Y\subset X$ is called cluster if $Y$ contains
at least two elements and for any points $a,b\in Y$ and $c\in X\setminus Y$
one has $\rho(a,b)<\rho (b,c)$.

\end{defn}

\begin{defn} \label{epsilon-clusters}
Let us fix a positive number $\varepsilon<1$. We say that
$Y$ is an $\varepsilon$-cluster if for any $c\in X\setminus Y$
one has $diam(Y)<\varepsilon \rho(c,Y)$, where $diam(Y)$ is the diameter
of the set $Y$.

\end{defn}
Clearly any $\varepsilon$-cluster is a cluster.
Clusters enjoy the following ``non-archimedean" property.

\begin{lmm} \label{intersection of two clusters}
If two clusters intersect non-trivially then one
of them contains the other one.

\end{lmm}

{\it Proof}. Let $Y_1$ and $Y_2$ be the clusters which intersect non-trivially.
Then we can choose a common element $c$. Let $a_1\in Y_1\setminus Y_2$ and
$a_2\in Y_2\setminus Y_1$. Then $\rho(a_1,c)<\rho(a_2,c)$, because $c\in Y_1$.
Since $c\in Y_2$ we have $\rho(a_2,c)<\rho(a_1,c)$. Contradiction. The lemma follows.
$\blacksquare$

Having a set $X$ with a metric $\rho$, as above, one can construct a tree $T$
in the following way. Tails of $T$ are parametrized by $X$.
Internal vertices are clusters. Two internal vertices are connected by
an edge if one cluster belongs to the other one. A tail vertex
corresponding to $x\in X$ is connected
to an internal vertex corresponding to a cluster $Y$ if $x\in Y$.

Let us fix a set $X$, and consider all metrics on $X$ such that
$diam(X)=1$. To every such a metric $\rho$ we can assign the tree
$T=T(X,\rho)$, as above.

The following proposition is easy to prove.

\begin{prp} \label{weights assigned to  vertices}
Let
 $\rho$ be a metric on $X$, as above.
One can assign numbers $\lambda_v, 0<\lambda_v<1$ to
all internal vertices $v$ of $T(X,\rho)$ in such a way
that for any two points $x_i,x_j\in X$ one has

$$C_1<{\rho(x_i,x_j)\over {\prod_{v<i,j}\lambda_v}}<C_2.$$

Here $C_i,i=1,2$ are positive numbers depending on $X$
(not on the metric), and the notation $v<i$
means that there is a path along the edges of $T$
which starts at the root vertex, ends at the tail vertex $i$
and contains $v$ (i.e. $v$ is ``closer" to the root than $i$).
\end{prp}

Let us fix the tree $T$
corresponding to a finite set $X$.
By definition $T$ has $|X|:=n$ tail vertices. Let us fix numbers $\lambda_v, v\in V_i(T)$
such that $0<\lambda_v<1$, the metric $\rho$ as in the Proposition 1,
and real numbers $\Delta, \Delta_1,...,\Delta_n$.
For a given subset $S\subset \{1,...,n\}$ we denote by $B_S$
the subset $x_i, i\in S$.

\begin{prp} \label{formula with weights}
The following formula holds:

$$(diam\,X)^{-\Delta}\prod_{v\in V_i(T)}\lambda_v^{\sum_{v<j}\Delta_j}=
C\,(diam\,X)^{-\Delta}\prod_{S\subset \{1,...,n\},|S|\ge 2}
(diam\,B_S)^{(-1)^{|S|+1}(\sum_{j\in S}\Delta_j)},$$

where $C$ is a positive number depending on the set $X$
(not on the metric).

\end{prp}

{\it Proof.} Straightforward. $\blacksquare$

We will denote the LHS of the formula from  Proposition \ref{formula with weights}
by $R(x_1,...,x_n,y)$ or by $R(x_1,...,x_n,y;\Delta_1,...,\Delta_n,\Delta)$.
In this notation
$x_1,...,x_n\in X$ corresponds to the tail
vertices of $T$ and $y$ corresponds to the root vertex.
Although the point $y$ is not an element of the set $X$,
one can imagine that all points $x_1,...,x_n,y$ belong
to a metric space $\widehat{X}=X\cup \{y\}$ such that
$X$ is a metric subspace of $\widehat{X}$, and
$\rho(y,x_i)=diam\,X$ for $1\le i\le n$.
The meaning of this notation will become clear later,
when we discuss the operator product expansion (abbreviated as OPE).
At this time we would like to make few comments
about the meaning of the function $R$.

 The function $R(x_1,...,x_n,y)$ will play the following
role in our considerations. Suppose we have local fields $\phi_1,...,\phi_n$
sitting at the points $x_1,...,x_n$ and having dimensions $\Delta_1,...,\Delta_n$.
When all points $x_i$ approach to the same point $y$, we can write
the OPE for the given fields:

$$\phi_1(x_1)...\phi_n(x_n)=\sum_{\lambda}
C_{\lambda}(x_1,...,x_n,y)\phi_{\lambda}(y),$$
where the coefficients $C_j(x_1,...,x_n,y)$ depend on the configuration
$\{x_1,...,x_n,y\}$ as well as on the dimensions $\Delta_{\lambda}$
of the fields $\phi_{\lambda}(y)$.

Let us take a local field $\phi=\phi_{\lambda}$
of dimension $\Delta$ sitting at $y$ such that $\phi$ appears in the OPE.
The function $R(x_1,...,x_n,y;\Delta_1,...,\Delta_n,\Delta)$
will be responsible for the singular part of the OPE. The tree $T$
appears because the points $x_i, 1\le i\le n$ can approach $y$
with different speeds. Geometrically we have a structure of cluster on
the set $\{x_1,...,x_n\}$. The point $y$ corresponds to the root of $T$.

We can describe the same picture in a slightly different way.
The correlator
$\lan \phi_1(x_1)...\phi_n(x_n)\ran$ has different behavior near
different strata of the Fulton-Macpherson compactification
of $Conf_n(X)$, where $X$ is the space-time. If the stratum $D_T$
corresponds to a tree $T$, then there are scaling factors $\lambda_v,
v\in V_i(T)$ such that the leading asymptotic
term of the correlator behaves near $D_T$
as the LHS of the formula from the Proposition \ref{formula with weights}  (up to a positive scalar).
This will be one of the axioms (which can be checked in all known examples).
Therefore the function $R$ represents the leading singular asymptotic
term of the correlator $\lan \phi_1(x_1)...\phi_n(x_n)\phi(y)\ran$.
 The constant factor $C$ depends
on the dimensions $\Delta, \Delta_i, 1\le i\le n$ as well as on
the tree $T$. One can think of it as of  function
which is bounded on any compact subset of the stratum $D_T$.

\begin{prp} \label{properties of function R}
The function $R$ satisfies the following properties:

1) If $n=0,1$ then $R=1$.

2) If $n\ge 2$ then

$$R=C_1\,(diam\,X)^{-\Delta}
\prod_{1\le i\le n}(min_{j\ne i, 1\le j\le n}\rho(x_i,x_j))^{-\Delta_i}$$
$$=C_2\,(diam\,X)^{-\Delta}min_{r_1,...,r_n}
\prod_{1\le i\le n}r_i^{-\Delta_i}.$$

Here $C_i, i=1,2$ are positive constants, and the second product is taken
over all disjoint union of balls
$\sqcup_{1\le i\le n} B(x_i,r_i)\subset X$.

\end{prp}

{\it Proof.} Notice that the original formula
for the function $R$ can be written as
$R=(diam\,X)^{-\Delta}
\prod_{1\le i\le n}(diam B_i)^{-\Delta_i},$ where $B_i$
is the diameter of the minimal cluster containing the point $x_i$.
This leads to a proof of the first equality. In order to prove
the second one, one can take $r_i={1\over 2}\rho_i, 1\le i\le n,$
where $\rho_i$ is the distance to a point closest to $x_i$.
$\blacksquare$

\subsection{Operator product expansion and QFTs on metric spaces}\label{OPE}

Let $(X,\rho)$ be a metric space of finite diameter. It will
play a role of the space-time.
We assume that we are given a family $H_x, x\in X$ of vector spaces,
each space is equipped with an increasing discrete filtration:

$$H_x=\bigcup_{\Delta\ge 0}H_x^{\le \Delta}.$$
We assume that each $H_x^{\le \Delta}$ is a finite-dimensional vector space,
which carries a norm $|\bullet|_x$.
We will call elements of $H_x^{\le \Delta}$ local fields at $x$
of dimension less or equal than $\Delta$.
In what follows we will assume that the filtrations are locally constant
(although this condition as well as many others can be relaxed).

Let $Conf_n(X), n\ge 0$ be the configuration space of $X$.
By definition $Conf_n(X)$ consists of sequences of $n$ pairwise
distinct points of $X$, and $Conf_0(X)=\emptyset$.

\begin{defn} \label{definition of OPE}
Let us fix $(x_1,...,x_n)\in Conf_n(X), y\in X$ and
non-negative numbers $\Delta_1,...,\Delta_n, \Delta$.

Operator product expansion (OPE) associated to these data is
a class of equivalence of
linear maps $m_n=m_n(x_1,...x_n,y;\Delta_1,...,\Delta_n,\Delta)$
such that

$$m_n:\otimes_{1\le i \le n}H_{x_i}^{\le \Delta_i}\to H_y^{\le \Delta},$$

and $m_n, n\ge 1$ satify the axioms listed below.
For $n=0$ it is the equivalence class of linear maps
$m_0:{\C}\to H_y^{\le \Delta}$.

\end{defn}

The equivalence relation is the following one.

\begin{defn} \label{equivalence relation}
Two maps $m_n$ and $m_n^{\prime}$ as above
are said to be equivalent if there exists $\varepsilon >0$
such that one has the following inequality for the norm of
the linear map:

$||m_n-m_n^{\prime}||\le C\,min \{r^{\Delta +\varepsilon}
\prod_{1\le i\le n}r_i^{-\Delta_i}\},$

where $C>0$ depends on the points $x_1,...,x_n$ (not on the metric
or dimensions), and the minimum is taken over all disjoint unions
of balls $\sqcup_{i=1}^{i=n}B(x_i,r_i)\subset B(y,r)$.

\end{defn}

Now we are going to list the axioms for $m_n, n\ge 0$ (or rather
properties of these maps):

{\it A1} (the norm inequality). One has:

$||m_n||\le C(\Delta_1,...,\Delta_n, \Delta)R(x_1,...,x_n,y)$,
where in the LHS we take the norm of any representative from the equivalence
class, and $R$ is the function introduced in the previous section.

{\it A2}. All maps $m_n$ from a given equivalence class are $S_n$-equivariant.

{\it A3}. Let $\Delta_i^{\prime}\le \Delta_i, 1\le i\le n$ and
$\Delta^{\prime\prime}\ge \Delta$ be fixed. Then the linear maps
$m_n$ and $j_2m_n^{\prime}j_1$, where $j_l, l=1,2$ are natural embeddings
associated with the filtrations,
are equivalent in the sense of the definition above (they are considered
as maps
$\otimes_{1\le i \le n}H_{x_i}^{\le \Delta_i^{\prime}}\to
H_y^{\le \Delta^{\prime\prime}}$).

{\it A4}. Let $n=1$, $\Delta\ge \Delta_1$ and $x_1=y$. Then the well-defined map
$m_1:H_y^{\le \Delta_1}\to  H_y^{\le \Delta}$ coincides with the natural
embedding induced by the filtration.

{\it A5} (operadic composition).
 For every planar tree $T$ with $n$ tails, and any choice of representatives
of $m_k$ assigned to internal vertices $v$ of $T$ in such a way
that $k$ is the incoming valency of $v$, the corresponding composition
map is equivalent to $m_n$ (for any choice of a point in the configuration space
and any choice of dimensions).

\begin{rmk}\label{equivalence classes}
1) It is easy to see that differences $m_n-m_n^{\prime}$ of equivalent maps
form a vector space. Hence the equivalence classes form a quotient
vector space.

2) Our definition of equivalent maps means that we do not consider
the input to OPE
of local fields of dimensions greater than $\Delta$.

\end{rmk}

\begin{que}\label{questions}
Suppose that $X$ is a Riemannian manifold.

1) Is it true that $m_1$ automatically gives a $D$-module structure
on the space of local fields $H$?

2) Can one treat fractional powers in the asymptotic series
for correlators in terms of asymptotic expansions at zero of certain
curves $f(t)$ in the moduli space of QFTs?

3) What is the deformation theory for $m_1$? First order
deformations are given by the elements of $H\otimes \Omega^d$.

\end{que}

\subsection{Quantum metric-measure spaces}\label{quantum metric-measure spaces}

Here we speculate about a possibility to generalize quantum Riemannian spaces discussed in the main body of the paper  to more general {\it quantum metric-measure spaces}.

Let us consider the following category ${\cal C}$. Objects of ${\cal C}$
are sequences $M:=((\varepsilon_1, Y_1),...,(\varepsilon_n, Y_n))$ where $\varepsilon_i, 1\le i\le n$
are postive numbers and $Y_i$ are compact metric spaces (as before, we will often denote  by $X$ the metric-measure space $(X,d_X)$) .
Let $M^{\prime}=((\varepsilon_1^{\prime}, Y_1^{\prime}),...,(\varepsilon_n^{\prime}, Y_m^{\prime}))$
be another object of ${\cal C}$.
A morphism $f:M\to M^{\prime}$ is defined by: 

1) a sequence 
$(X,X_1^{\varepsilon_1},...X_n^{\varepsilon_n}, X_{n+1}^{\varepsilon_1^{\prime}},...,X_{n+m}^{\varepsilon_m^{\prime}})$, where $X$ is a compact
metric space, $X_k^{\varepsilon_k},X_{l}^{\varepsilon_l^{\prime}}$ are non-intersecting
compact metric subspaces;

2) isometry embeddings $i_k:Y_k\to X, i_k^{\prime}:Y_k^{\prime}\to X$ such that

the closure of the $\varepsilon_k$-neighborhood of $i_k(Y_k)$ is equal to  $X_k^{\varepsilon_k}$,
and similarly for $i_k(Y_k^{\prime})$ and $X_{k+n}^{\varepsilon_k^{\prime}}, 1\le k\le m$.

Let us call such a morphism a {\it metric bordism between $M$ and $M^{\prime}$}. Let us formally add the empty metric space to ${\cal C}$. Then the metric bordism between $M$ and the empty space is
a sequence of isometric embeddings $i_k$ as above. Let us call the corresponding $W$ the {\it metric collar}
of $M$.
Composition of morphisms is defined in the following way.
Let $W_1$ and $W_2$ be two metric bordisms representing morphisms
$f_1:M\to M^{\prime}$ and $f_2:M^{\prime}\to M^{\prime\prime}$. Then we can construct a metric bordism
$W$ between $M$ and $M^{\prime}$ representing the composition $f_2\circ f_1$ such as follows. As a topological space $W$ is obtained from $W_1$ and $W_2$ by the gluing along canonically isometrically identified  compact subsets $X_{k+n}^{\varepsilon_k^{\prime}}, 1\le k\le m$ which belong to both
$W_1$ and $W_2$. Let $j_k$ denotes this isometric identification. The distance between $w_1\in W_1$ and $w_2\in W_2$ is defined
in the following way:

a) if both $w_1$ and $w_2$ belong to one of $X_{k+n}^{\varepsilon_k^{\prime}}$ then $d_W(w_1,w_2)$
is the distance inside of $X_{k+n}^{\varepsilon_k^{\prime}}$;

b) otherwise we define $d_W(w_1,w_2)$ as a minimum of the numbers $d_{W_1}(w_1,y)+d_{W_2}(j_k(y),w_2)$
where the minimum is taken over all points $y$ belonging to the union of the subsets
$X_{k+n}^{\varepsilon_k^{\prime}}$.

One checks that the distance function $d_W$ is symmetric and satisfies the triangle inequality.

The category ${\cal C}$ carries a symmetric monoidal structure with the tensor product
given by the disjoint union of ordered sequences, e.g. 
$(\varepsilon_1, Y_1)\otimes (\varepsilon_2, Y_2)=((\varepsilon_1, Y_1),(\varepsilon_2, Y_2))$.

%FIGURE 2 (composition of metric bordisms)

\begin{figure}[h!]
%\centerline{{Figure-2-metric-bordisms.pdf}}
\centering
\scalebox{0.5}
{\includegraphics{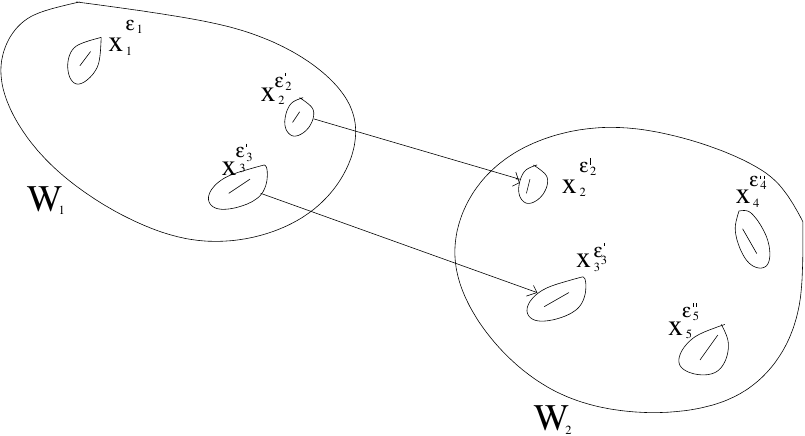}}
\end{figure}

\begin{defn} \label{quantum space as a functor}
A quantum metric space is a monoidal functor $F:{\cal C}_X\to Hilb_{\C}$.

\end{defn}

Let $W$ be a metric collar for an object $M$ as above. Then $W$ is also a metric collar for
all $M_{\delta_1,...,\delta_n}$ where $\delta_j<\varepsilon_j, 1\le j\le n$ and the metric spaces
$Y_j, 1\le j\le n$ are the same. We say that $M$ and $M_{\delta_1,...,\delta_n}$ are equivalent in $W$.
Suppose that all metric spaces above are in fact metric-measure spaces. Then we have the following
version of the above category. For a fixed metric collar $W$ of $M$ we fix 
$\varepsilon_2,\varepsilon_3,...,\varepsilon_n$ and let $\varepsilon:=\varepsilon_1\to 0$.
Suppose that the measure $d\mu_X$ being restricted to $X_1^{\varepsilon}$ admits an
asymptoric expansion $d\mu_X=\varepsilon^{l_1}d\mu_{Y_1}+o(\varepsilon^{l_1})$, where $l_1\ge 0$ and similarly for other
$Y_k$. We obtain a sequence $(l_1,...,l_n)$ of non-negative real numbers which we call exponents
of $M$ with respect to $W$. Then we define the category ${\cal C}^{mes}$ with objects which are equivalence
classes as above,
and in addition we assume that  all $Y_k, Y_j^{\prime}, X$ are metric-measure spaces. In the definition
of a morphism we will require that the bordism between $M$ and $M^{\prime}$ satisfies the above-mentioned
property for the measures. One checks that in this way ${\cal C}^{mes}$ becomes a symmetric monoidal
category. There is a natural monoidal functor $F:{\cal C}^{mes}\to Hilb_{\C}$ such that
$F((\varepsilon_1, Y_1),...,(\varepsilon_n, Y_n))=\otimes_{1\le i\le n}L_2(Y_i,d\mu_i)$.

\vspace{3mm}

{\bf References}

\vspace{3mm}

[AGS] L. Ambrosio, N. Gigli, G. Savare,
Gradient Flows: In Metric Spaces and in the Space of Probability Measures,
Birkhauser, 2005.

\vspace{2mm}

[BBG] P. Berard, G. Besson, S. Gallot, Embedding Riemannian manifolds
by their heat kernel, Geom. Funct. Anal., 4:4, 1994, 373-398.

\vspace{2mm}

[Ba] D. Bakry, Functional inequalities for Markov semigroups,
preprint, available at:
http://www.lsp.ups-tlse.fr/Bakry/

\vspace{2mm}

[BaEm] D. Bakry, M. \'Emery, Diffusions hypercontractives, Lect. Notes in Math. no. 1123, 1985, 177-206.

\vspace{2mm}

[BiVo] P. Biane, D. Voiculescu, A Free Probability Analogue of the Wasserstein Metric on the Trace-State Space, math.OA/0006044.

\vspace{2mm}

[ChC1] Cheeger, T.H. Colding, On the structure of spaces with Ricci curvature bounded below I,
J. Diff. Geom., 46, 1997, 37-74.

\vspace{2mm} 

[ChC2] Cheeger, T.H. Colding, On the structure of spaces with Ricci curvature bounded below I,
J. Diff. Geom., 54:1, 2000, 13-35.

\vspace{2mm}

[ChC3]J. Cheeger, T.H. Colding, On the structure of spaces with Ricci curvature bounded below III,
J. Diff. Geom., 54:1, 2000, 37-74.

\vspace{2mm}

[Co1] A. Connes, Non-commutative geometry, Academic Press, 1994.

\vspace{2mm}

[CoKr] A. Connes, D. Kreimer, Renormalization in quantum field theory and the Riemann-Hilbert problem, hep-th/9909126.

\vspace{2mm}

[CoMar] A. Connes, M. Marcolli, A walk in the non-commutative garden,
math.QA/0601054.

\vspace{2mm}

[Dou 1] M. Douglas, The statistics of string/M theory vacua, hep-th/0303194.

\vspace{2mm}

[Dou 2] M. Douglas, Talk at the String-2005 Conference,

http://www.fields.utoronto.ca/audio/05-06/strings/douglas.

\vspace{2mm}

[Dou3] M. Douglas, Spaces of Quantum Field Theories, arXiv:1005.2779.

\vspace{2mm}

[Dou L] M. Douglas, Z. Lu, Finiteness of volume of moduli spaces, hep-th/0509224.

\vspace{2mm} 

[En] A. Engoulatov, Heat kernel and applications to the convergence of Graph Field Theories,
preprint, 2006.

\vspace{2mm}

[FFRS] J. Fjelstad, J. Fuchs, I. Runkel, C. Schweigert, Topological and conformal field theory as Frobenius
algebras, math.CT/0512076.

\vspace{2mm}

[FG] J. Fr\"olich, K. Gawedzki, Conformal Field Theory and
geometry of strings, hep-th/9310187.

\vspace{2mm}

[FM] W. Fulton, R. Macpherson, A compactification of configuration spaces,
Annals Math., 139(1994), 183-225.

\vspace{2mm}

[Fu1] K. Fukaya, Collapsing of Riemannian manifolds and eigenvalues of Laplace
operator, Invent. Math., 87, 1987, 517-547.

\vspace{2mm}

[Gaw] K. Gawedzki, Lectures on Conformal Field Theory, in: Quantum Fields
and Strings: a course for mathematicians, AMS,1999, vol. 2,  727-805.
\vspace{2mm}

\vspace{2mm}

[Gi1] V. Ginzburg, Lectures on Noncommutative Geometry, math.AG/0506603.

\vspace{2mm}
[GiKa] V. Ginzburg, M. Kapranov, Koszul duality for operads,
Duke Math. J. 76 (1994), 203-272.
\vspace{2mm}

[Gro1] M. Gromov, Metric structures for Riemannian and non-Riemannian
spaces, Birkh\"auser, 1999.

\vspace{2mm}

[Gro2] M. Gromov, Random walks in random groups, Geom. Funct. Anal. 13:1, 2003, 73-146.

\vspace{2mm}

[Kaw 1] Y. Kawahigashi, Classification of operator algebraic conformal field theories in dimensions one and two, math-ph/0308029.

\vspace{2mm}

[Kaw 2] Y. Kawahigashi, Classification of operator algebraic conformal field theories,
math.OA/0211141.

\vspace{2mm}

[Kaw-Lo] Y. Kawahigashi, R. Longo, Noncommutative Spectral Invariants and Black Hole Entropy,
math-ph/0405037.

\vspace{2mm}

[KS] K. Kuwae, T. Shioya, Convergence of spectral structures: a functional analytic theory and its applications to spectral geometry, Comm. Anal. Geom., 11:4, 2003, 599-673.

\vspace{2mm}

[KMS] K. Kuwae, Y. Machigashira, T. Shioya, Sobolev spaces, Laplacian and heat kernel on Alexandrov spaces, 1998.

\vspace{2mm}

[KaKu1] A. Kasue, H.Kumura, Spectral convergence of Riemannian manifolds, Tohoku Math. J., 46, 1994,
147-179.

\vspace{2mm}

[KaKu2] A. Kasue, H.Kumura, Spectral convergence of Riemannian manifolds, II Tohoku Math. J., 48.1996,
71-120.

\vspace{2mm}

[Kok] S. Kokkendorff,  A Laplacian on metric measure spaces. Preprint of
Technical University of Denmark, March 2006.

\vspace{2mm}

[KoSe] M. Kontsevich, G.Segal, Wick rotation and the positivity of energy in quantum field theory, arXiv:2105.10161.
\vspace{2mm}

[KoSo1] M. Kontsevich, Y. Soibelman, Homological Mirror Symmetry and
torus fibrations, math.SG/0011041.

\vspace{2mm}

[KoSo2] M. Kontsevich, Y. Soibelman, Deformations of algebras over operads and
Deligne conjecture, math.QA/0001151, published in Lett. Math. Phys. (2000).

\vspace{2mm}

[KoSo3] M. Kontsevich, Y. Soibelman, Deformation theory, (book
in preparation).
\vspace{2mm}

[Li] H. Li, C*-algebraic quantum Gromov-Hausdorff distance,
math.OA/0312003.

\vspace{2mm}

[L] J. Lott, Optimal transport and Ricci curvature for metric-measure spaces, math.DG/06101542.

\vspace{2mm}

[LV] J. Lott, C. Villani, Ricci curvature for metric-measure spaces via optimal transport, math.DG/0412127.

\vspace{2mm}

[Led] M. Ledoux, The geometry of Markov diffusion generators,
preprint, available at:
http://www.lsp.ups-tlse.fr/Ledoux/

\vspace{2mm}

[OVa] H. Ooguri, C. Vafa, On the geometry of the string landscape and the swampland, hep-th/0605264.

\vspace{2mm}

[OW] H. Ooguri, Y. Wang, Universal Bounds on CFT Distance Conjecture, arXiv:2405.00674.

\vspace{2mm}

[Rie] M. Rieffel, Gromov-Hausdorff Distance for Quantum Metric Spaces,
math.OA/0011063.

\vspace{2mm}

[RW] D. Roggenkamp, K. Wendland, Limits and degenerations of unitary
Conformal Field Theories, hep-th/0308143.

\vspace{2mm}

[Ru] I. Runkel, Algebra in braided tensor categories and conformal field theory, preprint.

\vspace{2mm}

[S] T. Shioya, Convergence of Alexandrov spaces and spectrum of Laplacian,
1998.

\vspace{2mm}

[Seg] G.Segal, The definition of Conformal Field Theory, in: Topology,
Geometry and Quantum Field Theory, Cambridge Univ. Press, 2004,
421-577.

\vspace{2mm}

[Si] L. Silberman, Addendum to ``Random walks on random groups" by M. Gromov,
Geom. Funct. Anal., 13:1, 2003.
\vspace{2mm}

[So1] Y. Soibelman, Collapsing conformal field theories, spaces with non-negative Ricci curvature and non-commutative geometry,
in: Mathematical foundations of quantum 
field theory and perturbative string theory, vol. 83 of Proc. Sympos. Pure Math., pp. 245–278, Amer. Math. Soc., Providence, RI, (2011).

\vspace{2mm}

[St] K-T. Sturm, On the geometry of metric measure spaces,
preprint 203, Bonn University, 2004.

\vspace{2mm}

[ST] S. Stolz, P. Teichner, Supersymmetric field theories and integral modular functions,
in preparation.

\vspace{2mm}

[T] D. Tamarkin, Formality of chain operad of small squares, math.QA/9809164.

\vspace{2mm}

[U] H. Urakawa, Convergence rates to equilibrium of the heat kernels on compact
Riemannian manifolds, preprint.

\vspace{2mm}

[V] C. Villani, Optimal transport, old and new, book in preparation.

\vspace{2mm}

[Va] C. Vafa, The string landscape and the swampland, hep-th/0509212.
\vspace{2mm}

[W] Wei Wu, Quantized Gromov-Hausdorff distance, math.OA/0503344.

\vspace{2mm} 

[Z] S. Zelditch, Counting string/M vacua, math-ph/0603066.

\vspace{3mm}

{\bf address: Yan Soibelman, Department of Mathematics, Kansas State University, Manhattan, KS 66506, USA

{email: soibel@math.ksu.edu}}

\end{document}